\documentclass[journal]{IEEEtran}

\usepackage[pdftex]{graphicx}
\graphicspath{{../pdf/}{../jpeg/}}
\DeclareGraphicsExtensions{.pdf,.jpeg,.png}
\usepackage{xcolor,soul,framed}
\usepackage[cmex10]{amsmath}
\usepackage{array}
\usepackage{mdwmath}
\usepackage{mdwtab}
\usepackage{eqparbox}
\usepackage{url}
\usepackage{cite}
\usepackage{amsmath,amssymb,amsfonts}
\usepackage{algorithm}
\usepackage{algorithmic}
\usepackage{graphicx}
\usepackage{textcomp}
\usepackage{booktabs}
\usepackage{multirow}
\usepackage{hyperref}
\usepackage{comment}
\usepackage{tikz}
\usetikzlibrary{positioning,arrows.meta}
\begin{document}
\bstctlcite{IEEEexample:BSTcontrol}
    \title{Amplitude-Belief Reinforcement Learning for Adaptive Cyber Defense in Partially Observable V2X Networks}

  \author{Anwar~Shah,~\IEEEmembership{Member,~IEEE,}
        Rohan~Farooq,~\IEEEmembership{Member,~IEEE,}
        Sajid~Anwer,
      Tallha Akram, 
      Usman Ghous,
      Sajid Ullah Khan\\
\thanks{Anwar Shah is with the Department of Data Science and Artificial Intelligence and Cyberarian Research Lab(email: anwar.shah@nu.edu.pk).}%
\thanks{Rohan Farooq is with Department of Computer Science (e-mail: rohan.farooq@nu.edu.pk).}
\thanks{Sajid Anwer is with Department of Software Engineering, College of Computer Engineering and Sciences, Prince Sattam Bin Abdulaziz University, Al-Kharj(email: s.anwer@psau.edu.sa).}%

   \thanks{Tallha Akram is with Department of Information Systems, College of Computer Engineering and Sciences, Prince Sattam Bin Abdulaziz University, Al-Kharj(email: t.akram@psau.edu.sa).}%
    \thanks{Usman Ghous is with Department of Computer Science(email: usman.ghous@nu.edu.pk).}%
   \thanks{Sajid Ullah Khan is with Department of Information Systems, College of Computer Engineering and Sciences, Prince Sattam Bin Abdulaziz University, Al-Kharj(email: sk.khan@psau.edu.sa).}%
   }


\maketitle

\begin{abstract}
The Internet of Vehicles (IoV) creates a partially observable and adversarial V2X communication environment in which malicious vehicles may probe, attack, and evade defensive mechanisms over time. Existing IoV intrusion-detection methods are often evaluated as static classification systems and therefore provide limited support for sequential mitigation under adaptive attacker behavior. This paper formulates IoV cyber defense as a partially observable sequential decision problem and proposes Quantum Belief-Integrated Reinforcement Defense (Q-BIRD), an amplitude-belief reinforcement learning framework for adaptive V2X defense. Q-BIRD represents uncertainty over hidden attacker intent through a normalized complex-valued belief state and converts amplitudes into intent probabilities through a Born-rule-inspired mapping. The resulting belief features are used by a Proximal Policy Optimization defender to select cost-aware mitigation actions, including monitoring, alerting, throttling, and isolation. Experiments are conducted in a SUMO--OMNeT++/Veins V2X co-simulation environment using IEEE 802.11p/DSRC-based V2V and V2I communication. Across 10 independent random seeds and 80 test episodes per seed, Q-BIRD reduces mean cumulative damage from \(36.0 \pm 5.5\) to \(28.0 \pm 3.0\) compared with PPO using classical Bayesian belief, corresponding to a 22.2\% reduction. It also reduces damage variance from \(12.0 \pm 2.8\) to \(6.0 \pm 1.5\), corresponding to a 50.0\% reduction. The attack success rate decreases from \(0.12 \pm 0.03\) to \(0.05 \pm 0.02\), while survival probability increases from \(0.91 \pm 0.02\) to \(0.96 \pm 0.02\). Communication-level results show that Q-BIRD maintains a packet delivery ratio of \(0.94 \pm 0.02\), latency of \(45 \pm 6\) ms, throughput of \(3.60 \pm 0.15\) Mbps, and service availability of \(0.95 \pm 0.02\). Explainability analysis using SHAP, LIME, and Grad-CAM suggests that belief-related features contribute strongly to mitigation decisions during attacker strategy transitions. These results indicate that amplitude-based belief modeling can improve both cyber-defense stability and V2X communication reliability under partial observability.
\end{abstract}

\begin{IEEEkeywords}
Internet of Vehicles, V2X security, reinforcement learning, partial observability, amplitude belief, adaptive cyber defense
\end{IEEEkeywords}

%
\IEEEpeerreviewmaketitle


\section{Introduction}

\noindent The Internet of Vehicles (IoV) enables large-scale connectivity among vehicles, roadside units, and cloud services to support safety-critical applications such as cooperative driving, traffic management, and autonomous transportation systems \cite{xu2023vehicular,li2024jsac}. While this connectivity improves efficiency and situational awareness, it also significantly expands the attack surface of vehicular networks, exposing IoV systems to cyber threats, including spoofing, denial-of-service (DOS), and coordinated intrusion attacks \cite{ullah2022iovsurvey}.

\noindent Current IoV intrusion detectors work in a fixed manner; they learn attack signatures from pre-recorded, labeled data and then apply those signatures in real-world traffic \cite{ring2022benchmarking}. In laboratory settings, the detectors achieve high accuracy because they treat attack patterns as constant and assume that opponents will not change their methods. Field studies demonstrate that this assumption fails once the system is in service, adversaries probe it, change their behavior, and evade the learned rules. The reliability of the detector falls rapidly \cite{sommer2010closedworld}.

\noindent Recent work seeks to overcome those limits by turning to reinforcement learning and game-theoretic models for cybersecurity. Those models frame network defense as a sequence of decisions between attackers and defenders, both acting strategically \cite{chen2022adversarial,zhang2023rlsurvey,durkota2022markov}. The defender repeatedly selects mitigations such as raising an alert, throttling traffic, or isolating a suspicious node while weighing detection success, operational cost, and service uptime. A primary difficulty is that IoV defense must proceed with only partial observation; the true intent of an attacker remains hidden, and visible signals are often ambiguous due to probing or evasion.

\noindent Classical probabilistic belief models, commonly used to manage uncertainty in partially observable environments, tend to collapse rapidly under noisy or deceptive observations, leading to overconfident decisions that can be exploited by adaptive adversaries \cite{ortega2022bounded,tschantz2023robust}. Recent advances in decision theory and learning under uncertainty suggest that non-classical uncertainty representations can improve robustness in environments characterized by ambiguity and delayed information \cite{zhang2023qirl}.

\noindent Although reinforcement learning-based defenses for the IoV have progressed, the existing approaches still rely on classical Bayesian belief models to track hidden attacker intent. These representations break down due to deceptive observations that adaptive attackers intentionally generate through probing and evasion. The outcome is an overconfident posterior that corrupts the policy input, triggering erratic, high-variance defensive actions at moments when reliable reasoning is most needed. Existing studies, such as Guo~et~al.~\cite{guo2025tvt} and
Chen~et~al.~\cite{chen2022adversarial} improve the policy- or game-theoretic aspects of IoV defense, but they treat the belief model as a fixed classical submodule rather than as a designable element of the defense architecture. 

This gap is significant: a defender whose belief collapses under evasion cannot maintain effective damage control, no matter how advanced the policy.
To overcome this challenge, non-classical uncertainty representations inspired by quantum probability theory preserve superposition across intent hypotheses under ambiguous observations without requiring quantum hardware, yet their application to IoV security remains entirely unexplored.

\noindent In this work, we propose a quantum-inspired belief representation for IoV defense that models uncertainty over hidden attacker intent using amplitude-based belief states. Unlike classical Bayesian belief updates, the proposed approach preserves uncertainty in the presence of ambiguous observations and enables more robust belief evolution without relying on quantum hardware or quantum computation \cite{zhang2023qirl}. We integrate this belief representation into a reinforcement learning-based defender using Proximal Policy Optimization (PPO) \cite{schulman2017ppo}, allowing the defender to select cost-aware mitigation actions in response to an adaptive attacker. The proposed approach has been evaluated in a simulated IoV security environment that explicitly models attacker adaptation through probing, attacking, and evasion strategies. Simulation-based evaluation enables a controlled analysis of long-term damage, stability, and robustness under adversarial behavior, which cannot be captured by offline datasets alone \cite{ring2022benchmarking}. Experimental results demonstrate that, compared to classical probabilistic belief modeling, the proposed quantum-inspired belief yields significantly lower variance in cumulative attack damage and more stable defensive behavior under shifts in the attacker's strategy, while maintaining comparable average performance. These results highlight the importance of robust uncertainty modeling for autonomous IoV security beyond static dataset-driven intrusion detection.

\noindent The main contributions of this work are as follows:
\begin{itemize}
    \item We formulate adaptive IoV cyber defense as a partially observable sequential decision problem in which the defender observes communication-derived risk features but does not directly observe the attacker's intent.
    \item We propose an amplitude-belief representation that maintains a normalized complex-valued belief state over hidden attacker modes and maps it to intent probabilities for policy conditioning.
    \item We integrate the proposed belief representation with a PPO-based defender that selects cost-aware V2X mitigation actions under security and service-availability constraints.
    \item We evaluate Q-BIRD in a SUMO--OMNeT++/Veins co-simulation environment using IEEE 802.11p/DSRC-based V2V and V2I communication.
    \item We compare Q-BIRD against no mitigation, rule-based IDS, DQN, DRQN, PPO without belief, PPO-LSTM, PPO with Bayesian belief, and moving-target defense baselines using both security-level and communication-level metrics.
    \item We analyze the trained policy using SHAP, LIME, and Grad-CAM to examine whether belief-related features contribute to mitigation decisions during attacker strategy transitions.
\end{itemize}

\noindent The remainder of this paper is organized as follows. Section~II reviews IoV intrusion detection, reinforcement learning-based defense, and belief modeling under uncertainty. Section~III formulates the partially observable IoV defense problem. Section~IV defines the threat model. Section~V presents the proposed Q-BIRD methodology. Section~VI reports the experimental setup, comparative results, ablation study, and explainability analysis. Section~VII concludes the paper and outlines future work.

\section{Literature Review}
The security of the Internet of Vehicles (IoV) has attracted significant research attention due to the safety critical nature of vehicular communication systems and their exposure to a wide range of cyber threats. Existing studies have primarily focused on intrusion detection mechanisms and preventive security strategies aimed at identifying malicious behavior within vehicular networks.

\subsection{IoV Intrusion Detection and Static Security Approaches}

\noindent Security for in vehicle networks and for the Internet of Vehicles has usually been provided by intrusion detection systems that use machine learning plus deep learning classifiers. A detailed survey by \cite{rajapaksha2022survey} and others shows that almost every current solution trains a supervised model on a static offline data set. Those systems reach high accuracy when the test traffic matches the training traffic but they treat both the traffic pattern but also the attacker as fixed once the attacker varies the pattern, accuracy falls.

\noindent Recent studies tries to raise detection scores improving the data itself. \cite{xu2025vehcom} supply extra synthetic samples so that their model recognises zero day attacks inside the Internet of Vehicles. \cite{khan2025srep} build a deep learning detector that is specialised for vehicular buses. Both upgrades still wait for an attack before they act they do not plan a sequence of defensive steps and they do not act in advance they cannot lower the cumulative harm that a patient attacker inflicts over time.

\subsection{Reinforcement Learning for Adaptive IoV Security}

\noindent Static intrusion detection has limits, and reinforcement learning is a way to let cyber defense adapt. Nguyen, besides Reddi \cite{nguyen2021tnnls} show that deep reinforcement learning fits security tasks where the setting changes and opponents act. Ju et al \cite{ju2023tits} use the same approach for vehicular networks, they protect task offloading at the vehicular edge and find that learned rules beat fixed plans when traffic patterns shift.

\noindent More recent work has explicitly modeled attacker  defender interactions in IoV systems. Guo et al. \cite{guo2025tvt} combine game theory and reinforcement learning to design moving target defense strategies for IoV, while Ren et al. \cite{ren2024iot} propose a multi agent deep reinforcement learning approach for autonomous security management in Internet of Things environments. Although these studies demonstrate the benefits of adaptive and sequential defense, they often assume that attacker behavior is either fully observable or accurately inferred, which is rarely achievable in practice.

\subsection{Risk Aware and Robust RL Based Security Models}

\noindent Given the high cost of security failures in vehicular systems, several works have incorporated risk awareness and robustness into reinforcement learning frameworks. Lu et al. \cite{lu2024iov} propose a risk aware federated reinforcement learning approach for secure IoV communications, emphasizing the trade off between system performance and security risk. Similarly, Liu et al. \cite{liu2023arxiv} analyze the robustness of deep reinforcement learning algorithms under false data injection attacks, demonstrating that learning based controllers can be vulnerable to deceptive inputs.

\noindent While these approaches move toward more realistic threat models, they rely primarily on classical probabilistic assumptions and do not explicitly address uncertainty in attacker intent. As a result, belief updates may become overconfident or unstable when observations are sparse, noisy, or adversarially manipulated.

\noindent Classical belief representations commonly used in reinforcement learning and game theoretic frameworks struggle under partial observability, particularly when attackers deliberately obfuscate their behavior. This limitation motivates the exploration of alternative belief representations that can preserve uncertainty and support robust decision making in early and ambiguous attack stages.

\subsection{Research Gap and Motivation}

\begin{figure}
    \centering
    \includegraphics[width=0.9\linewidth]{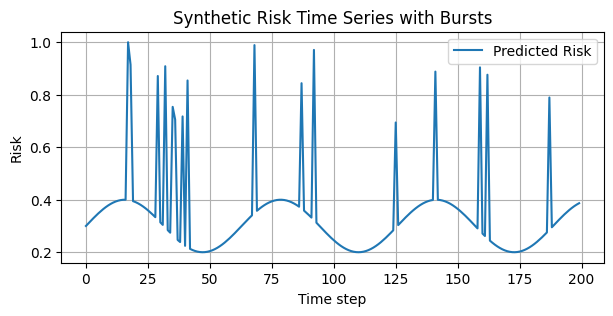}
    \caption{Risk signal with transient bursts showing why instantaneous classification fails in IoV environments.}
    \label{fig:risk}
\end{figure}

\noindent The above review shows that existing IoV security research either focuses on static intrusion detection or employs reinforcement learning with limited treatment of uncertainty in attacker intent. Although adaptive and risk aware reinforcement learning based defenses have been proposed, belief modeling remains underexplored, especially in environments with deceptive and non stationary adversaries. Fig.~\ref{fig:risk} illustrates non stationary and burst driven risk evolution in IoV environments. Transient high magnitude spikes correspond to probing or attack attempts, while smooth low amplitude variations represent benign traffic fluctuations. Such ambiguous signals make instantaneous classification unreliable and motivate belief based sequential defense.

\noindent Fig.~\ref{fig:risk} illustrates why single-step classification is insufficient for IoV defense. Short-lived bursts may correspond to probing or early-stage attacks, while low-amplitude variations may still carry delayed risk. This motivates sequential belief tracking rather than isolated packet-level classification.

\noindent In contrast to prior work, this study introduces a quantum inspired belief representation within a reinforcement learning based IoV defense framework. By modeling attacker intent uncertainty using amplitude based belief states rather than purely probabilistic updates, the proposed approach aims to improve robustness and stability under partial observability. This integration of belief dynamics into the defender’s decision making loop distinguishes the proposed method from existing intrusion detection, reinforcement learning, and game-theoretic approaches, and enables effective long-term damage mitigation in a sequential attacker defender setting.

\section{Problem Formulation}

\noindent We consider Internet of Vehicles (IoV) security as a sequential decision making problem under partial observability. At each time step $t$, the defender observes a system state $s_t$ but does not directly observe the attacker’s intent $\theta_t \in \Theta$. Instead, the defender must make decisions based on indirect, noisy, and potentially deceptive observations of system behavior.

\noindent The defender selects a mitigation action $a_t$, such as monitoring, throttling, or isolation. This action influences both the immediate damage incurred by the system and the evolution of future system states. The damage depends on the attacker’s true intent and the effectiveness of the chosen defensive action. Since the attacker’s intent is hidden, the defender cannot reliably minimize damage using instantaneous or myopic decisions.

\begin{figure}
    \centering
    \includegraphics[width=1\linewidth]{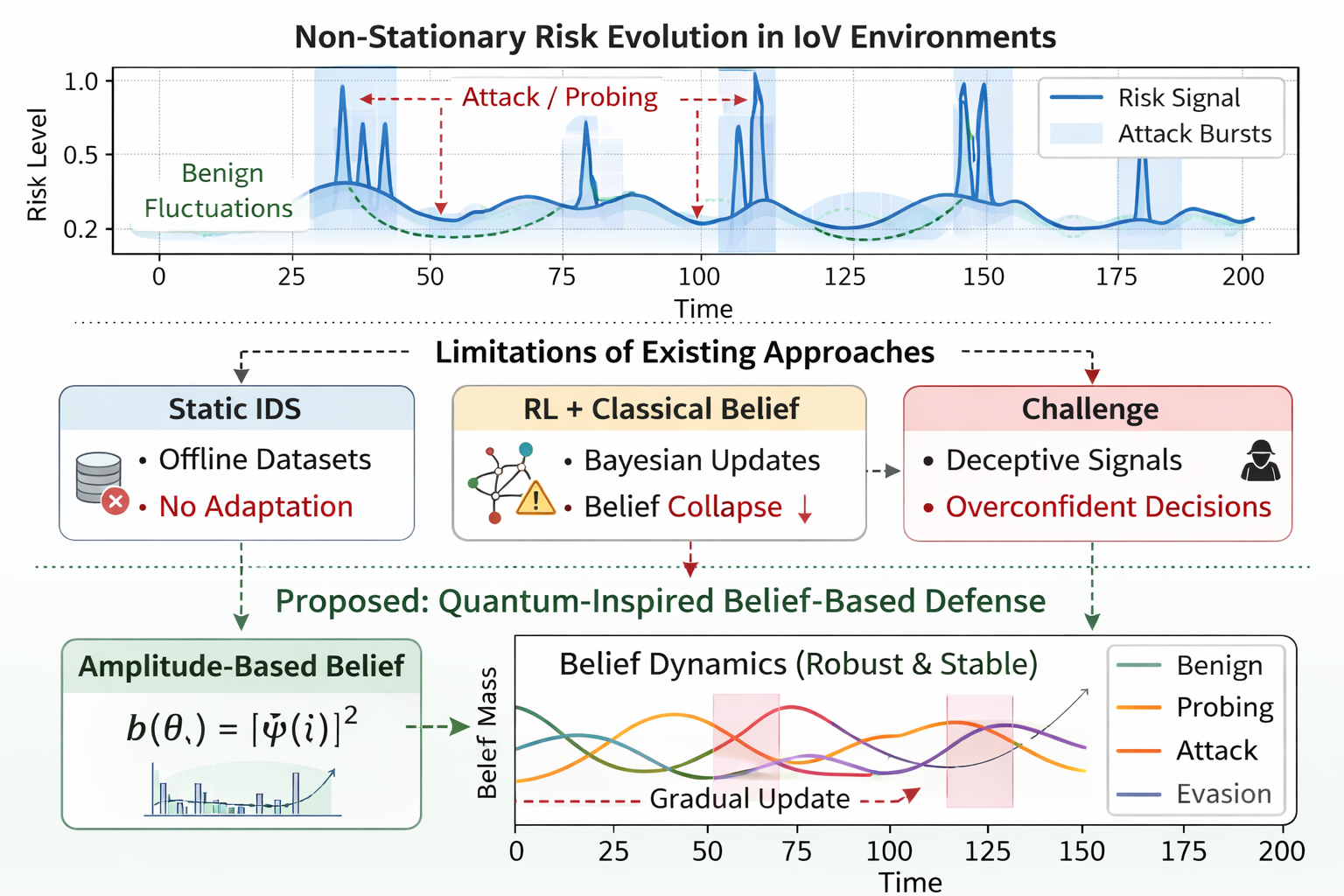}
    \caption{Non-stationary risk evolution under ambiguous IoV observations. Classical belief updates may concentrate rapidly under deceptive evidence, whereas amplitude-based belief evolution is designed to preserve uncertainty during attacker strategy transitions.}
    \label{fig:risk_evolution}
\end{figure}
\noindent Fig.~\ref{fig:risk_evolution} conceptually contrasts direct probabilistic belief concentration with amplitude-space belief evolution. The figure supports the problem formulation by showing that the defender must operate under delayed and ambiguous evidence rather than under fully observable attacker states.

\noindent The defender’s objective is therefore to minimize the expected cumulative damage while accounting for the operational cost of defensive actions:
\begin{equation}
\min \; \mathbb{E}\left[\sum_{t=1}^{T} \gamma^{t-1}
\left(D_t + C(a_t)\right)\right],
\end{equation}
where $D_t$ denotes the effective damage incurred at time step $t$, $C(a_t)$ represents the cost associated with the selected defense action, and $\gamma \in (0,1]$ is a discount factor that balances immediate and future impact. The expectation reflects uncertainty in attacker behavior, system dynamics, and observations.

\noindent This formulation emphasizes long-term damage mitigation under uncertainty, rather than instantaneous intrusion detection decisions, and serves as the basis for designing adaptive IoV defense policies.

\section{Threat Model}
\label{sec:threat}
We model an adaptive attacker capable of dynamically adjusting its behavior in response to defensive actions. The attacker keeps an internal intent value that remains invisible to outsiders, which can correspond to benign activity, probing, active attack, or evasion. The defender never sees the intent value itself but notices side effects that leak through events like anomalous traffic bursts, low-intensity probes, or
evasive behavior that lower the chance of discovery. 

\begin{figure}
    \centering
    \includegraphics[width=1\linewidth]{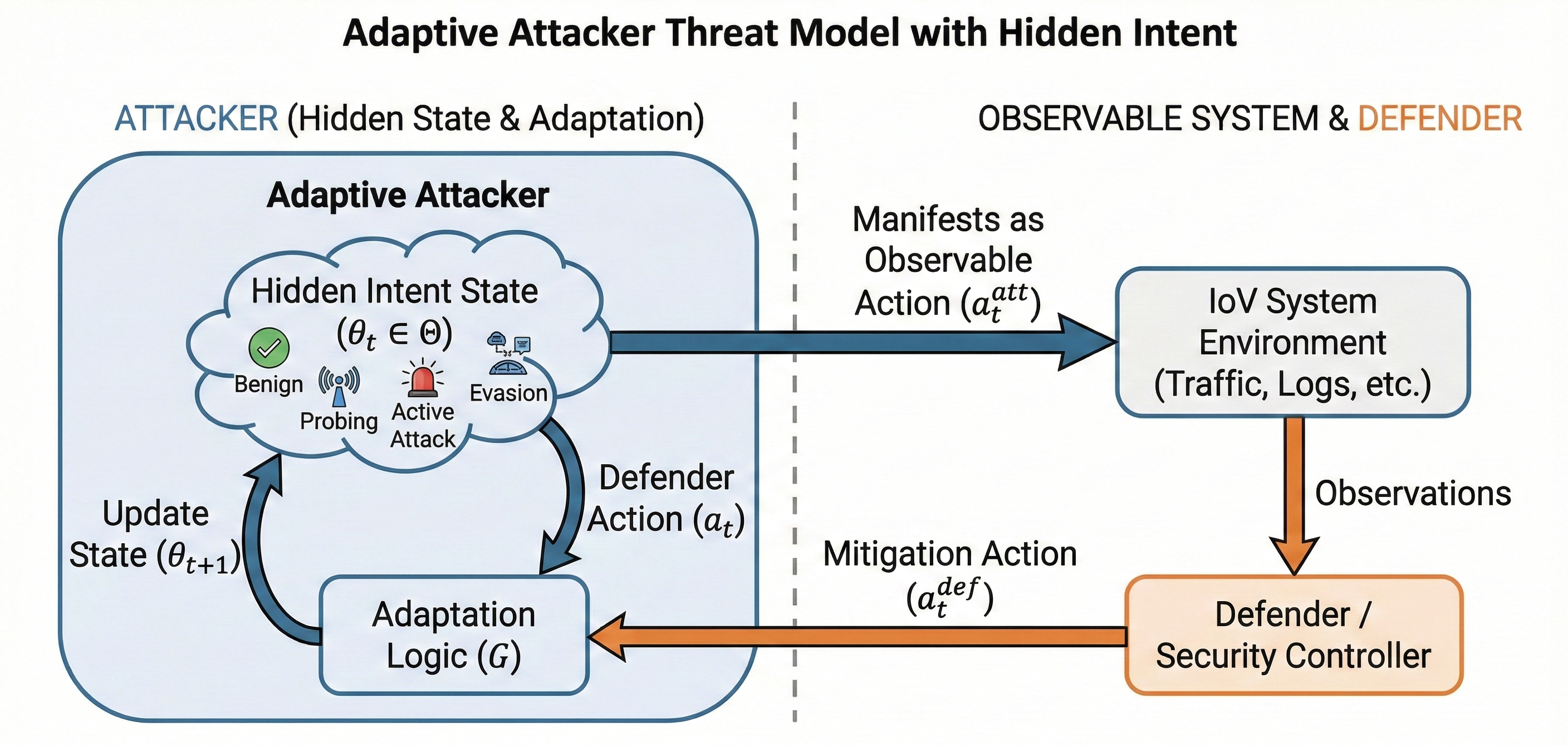}
    \caption{Attacker operates on hidden intent while the defender sees only system state and responds via policy $\pi_\phi$.}    
    \label{fig:threat_model}
\end{figure}
\noindent Fig.~\ref{fig:threat_model} summarizes the attacker--defender interaction assumed in the threat model. The attacker operates through a hidden intent state, while the defender observes only external communication and risk signals. This separation between hidden intent and observable evidence is the source of partial observability in the proposed formulation.

\noindent When a defender acts, the attacker changes how it behaves based on how much the system tries to find it. In this process, the attacker also considers how the defense reacted in the past. In vehicle networks, real actors operate in this way. It is also how they function in connected systems. By using this method, the model avoids the fixed patterns that datasets often show. Because of this design, the model shows how two sides influence each other. Such interactions are what researchers describe in recent studies.

\noindent Let the attacker intent space be
\begin{equation}
\Theta = \{\theta_1, \theta_2, \dots, \theta_H\}.
\end{equation}
At each time step, the attacker selects an action
\begin{equation}
a_t^{att} \in \mathcal{A}^{att},
\end{equation}
based on its current intent $\theta_t$. The attacker’s internal state evolves adaptively according to
\begin{equation}
\theta_{t+1} = G(\theta_t, a_t^{def}),
\end{equation}
where $G : \Theta  \times \mathcal{A}^{def} \rightarrow \Theta$ models adversarial adaptation in response to defensive actions. This transition captures realistic strategic behavior in which attackers escalate, probe, or evade depending on mitigation pressure. The inclusion of $G$ ensures that the environment is non stationary and adversarial, distinguishing this framework from static intrusion detection settings.

\section{Methodology}

\begin{figure}[h!]
    \centering
\includegraphics[width=1\linewidth]{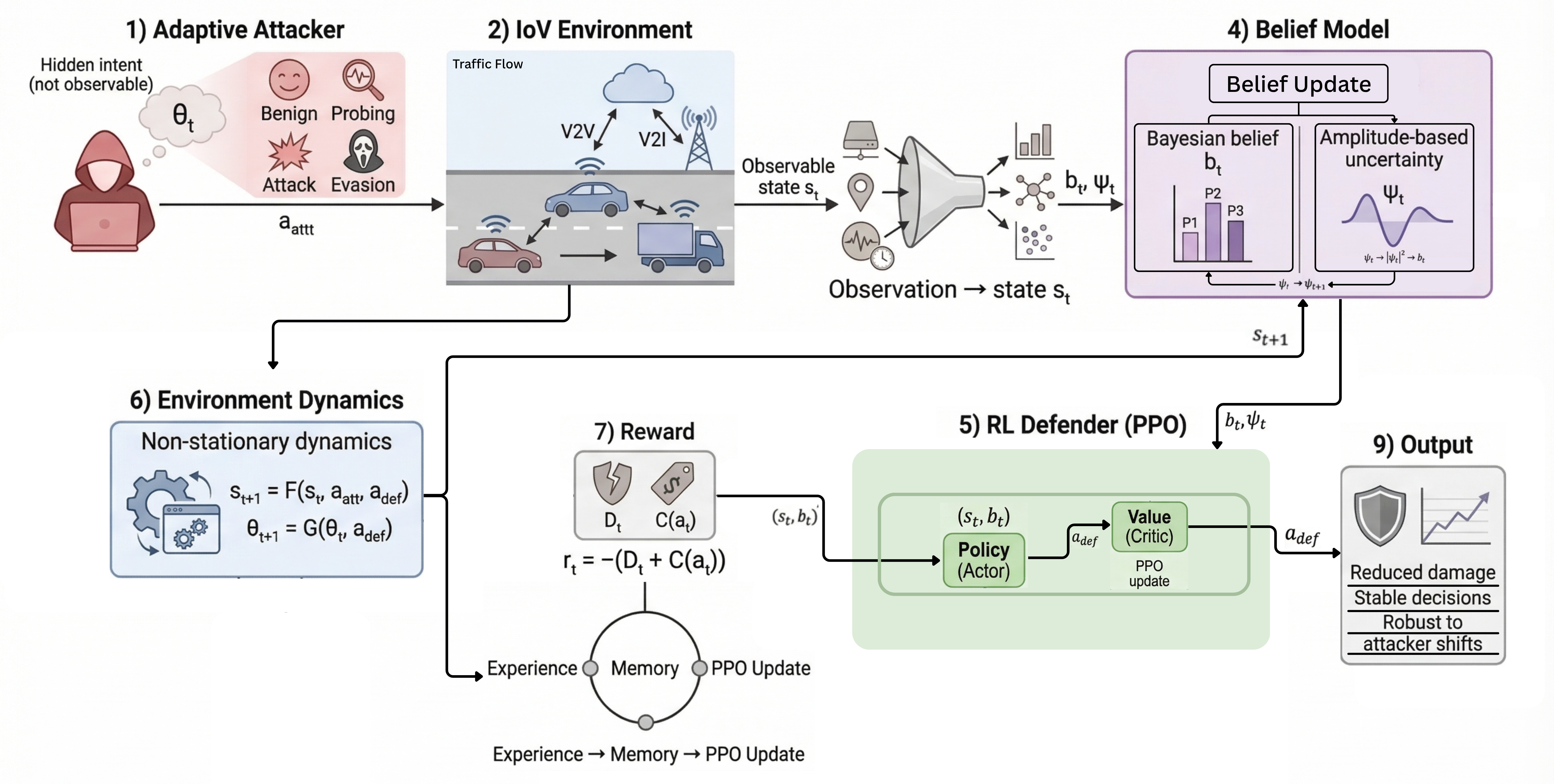}
    \caption{Methodological flow of Q-BIRD showing the interaction among observable IoV state, hidden attacker intent, amplitude-based belief update, PPO-based defense policy, mitigation action, and reward feedback.}
    \label{fig:Q-BIRD}
\end{figure}

\noindent This section provides a formal description of the proposed Internet of Vehicles (IoV) defense framework. The system is modeled as a sequential decision-making process under partial observability, involving interactions between an adaptive attacker and an autonomous defender over time. The formulation is illustrated graphically in Figure \ref{fig:Q-BIRD} to clearly define system dynamics, belief evolution, and policy optimization, independent of implementation-specific details.

\noindent Fig.~\ref{fig:Q-BIRD} shows the closed-loop structure of the proposed framework. The key distinction from a standard PPO defender is the belief-update module, which transforms ambiguous observations into a belief-conditioned policy input before action selection.
\subsection{Sequential Interaction Model}

\noindent We consider a discrete time decision process indexed by $t = 1,2,\dots,T$. At each time step, an adaptive attacker interacts with an autonomous defender within a dynamic IoV environment. The complete latent system state is defined as
\begin{equation}
x_t = (s_t, \theta_t),
\end{equation}
where $s_t \in \mathcal{S} \subseteq \mathbb{R}^n$ denotes the observable IoV system state and $\theta_t \in \Theta = \{\theta_1,\dots,\theta_H\}$ represents the hidden attacker intent. The observable state $s_t$ consists of aggregate traffic statistics, anomaly indicators, and risk related signals extracted from vehicular communications. In contrast, $\theta_t$ encodes the attacker’s internal strategic mode, such as benign behavior, probing, active attack, or evasion.

\noindent The defender observes $s_t$ but does not directly observe $\theta_t$, which introduces partial observability into the system. As a result, the defender must infer attacker intent indirectly from observable signals over time.

\subsection{Environment Transition Dynamics}

\noindent The observable IoV system evolves according to the transition function
\begin{equation}
s_{t+1} = F(s_t, a_t^{att}, a_t^{def}),
\end{equation}
where $F : \mathcal{S}  \times \mathcal{A}^{att}  \times \mathcal{A}^{def} \rightarrow \mathcal{S}$. This function captures the joint influence of attacker and defender actions on traffic flow, service availability, and risk signals. Defensive actions such as throttling or isolation may reduce attack impact but can also affect system performance. The complete hidden state evolution can therefore be expressed as
\begin{equation}
x_{t+1} = \big(F(s_t, a_t^{att}, a_t^{def}), \; G(\theta_t, a_t^{def})\big).
\label{eq:joint_dynamics}
\end{equation}
The above equation defines the joint evolution of the complete system state. Recall that the full latent state at time $t$ is given by $x_t = (s_t, \theta_t)$, where $s_t$ denotes the observable IoV system state and $\theta_t$ represents the hidden attacker intent. The expression~\ref{eq:joint_dynamics} therefore states that the next system state consists of two simultaneously updated components. The first component, $F(s_t, a_t^{att}, a_t^{def})$, determines the next observable state $s_{t+1}$ as a function of the current state, the attacker’s action $a_t^{att}$, and the defender’s action $a_t^{def}$. This captures how malicious activity and defensive interventions jointly influence traffic conditions, congestion levels, and risk related signals in the IoV environment. The second component, $G(\theta_t, a_t^{def})$, determines the next hidden attacker intent $\theta_{t+1}$ based on the current intent and the defender’s action. This models adversarial adaptation, whereby the attacker may modify its internal strategy in response to defensive pressure. Together, this formulation compactly represents the simultaneous evolution of both the observable environment and the hidden attacker strategy, thereby defining the complete system dynamics under partial observability.

\subsection{Belief Modeling Under Partial Observability}

\noindent As $\theta_t$ is not directly observable, the defender maintains an internal belief over the intent space. In the classical formulation, the belief state is represented as a probability vector
\begin{equation}
b_t \in \Delta^{H-1}
\end{equation}
which states that the belief at time $t$, denoted by $b_t$, belongs to the $(H-1)$ dimensional probability simplex. This means that $b_t$ is not an arbitrary vector but a valid probability distribution defined over the $H$ possible attacker intent states in $\Theta$. In other words, the defender maintains a structured probabilistic estimate over the hidden intent variable.

\noindent The probability simplex is formally defined as
\begin{equation}
\Delta^{H-1} = \left\{ b \in \mathbb{R}^H : \sum_{i=1}^{H} b_i = 1,\; b_i \ge 0 \right\}.
\end{equation}
This definition specifies that every belief vector $b$ is a collection of $H$ numbers. For each component $b_i$, the value is the likelihood that an attacker has intent $\theta_i$. By adding all the $b_i$ values, the total is always 1 ($\sum_{i=1}^{H} b_i = 1$). There is also a constraint that no $b_i$ is less than zero. To be a valid probability distribution, $b_t$ must follow those rules.

\noindent The belief state changes over time according to
\begin{equation}
b_{t+1} = \mathcal{B}(b_t, s_{t+1}, a_t^{def}),
\label{eq:bayesian_belief}
\end{equation}
where $\mathcal{B}$ is the belief update operator. The new belief $b_{t+1}$ is computed from the old belief $b_t$, from the system state $s_{t+1}$ that was just observed and from the defender action $a_t^{def}$. Operator $\mathcal{B}$ performs one Bayesian inference step, this revises the prior belief with evidence that the observable part of the environment supplies. By repeating this update at every step, the defender keeps improving its estimate of the attacker's hidden goal even when only partial observations are available.

\noindent To avoid premature concentration of belief under ambiguous observations, Q-BIRD represents uncertainty in an amplitude space. Let
\begin{equation}
\psi_t \in \mathbb{C}^{H}, \qquad \|\psi_t\|_2^2 = 1,
\end{equation}
where each component \(\psi_t(i)\) corresponds to attacker intent \(\theta_i\). The intent probability used by the defender is obtained through the Born-rule-inspired mapping
\begin{equation}
b_t(\theta_i)=|\psi_t(i)|^2, \qquad \sum_{i=1}^{H} b_t(\theta_i)=1.
\end{equation}

\noindent The amplitude state is updated using an observation- and action-conditioned transformation:
\begin{equation}
\tilde{\psi}_{t+1}=M(o_{t+1},a_t^{def})U(a_t^{def})\psi_t,
\label{eq:amplitude_evidence}
\end{equation}
\begin{equation}
\psi_{t+1}=\frac{\tilde{\psi}_{t+1}}{\|\tilde{\psi}_{t+1}\|_2+\epsilon_n}.
\label{eq:amplitude_update}
\end{equation}
where \(o_{t+1}\) denotes the observation extracted from the next IoV state, \(U(a_t^{def})\in \mathbb{C}^{H\times H}\) controls amplitude redistribution across intent hypotheses under the selected mitigation action, \(M(o_{t+1},a_t^{def})\) is a diagonal evidence operator that weights intent hypotheses according to observation compatibility, and \(\epsilon_n\) is a small numerical constant used to avoid division by zero.

\noindent In the implemented version, \(M(o_{t+1},a_t^{def})\) is defined as
\begin{equation}
\begin{aligned}
M(o_{t+1},a_t^{def}) &=
\mathrm{diag}\big(
\exp\!\left[-\frac{\kappa}{2}\ell_1(o_{t+1},a_t^{def})\right], \\
&\qquad\qquad
\dots,
\exp\!\left[-\frac{\kappa}{2}\ell_H(o_{t+1},a_t^{def})\right]
\big),
\end{aligned}
\end{equation}
where \(\ell_i(o_{t+1},a_t^{def})\) is the observation mismatch between the current IoV risk signal and the expected signature of intent \(\theta_i\), and \(\kappa>0\) controls evidence sharpness. This construction preserves uncertainty when the observation is ambiguous and concentrates probability mass only when repeated evidence consistently supports a specific intent.

\noindent The PPO policy receives the belief-conditioned input
\begin{equation}
z_t=[s_t \,\|\, b_t],
\end{equation}
where \(s_t\in\mathbb{R}^{8}\) is the observable IoV state and \(b_t\in\mathbb{R}^{4}\) is the probability vector obtained from \(\psi_t\). Thus, the reinforcement learning policy remains a classical neural policy, while the belief update uses an amplitude-space representation to regulate uncertainty under partial observability.

\subsection{Damage and Reward Modeling}

\noindent The instantaneous damage incurred at time $t$ is defined as
\begin{equation}
D_t = d(\theta_t, a_t^{def}),
\end{equation}
where $\theta_t \in \Theta$ denotes the true (but hidden) attacker intent at time $t$, and $a_t^{def} \in \mathcal{A}^{def}$ represents the defensive action selected by the defender. The function \[
d : \Theta  \times \mathcal{A}^{def} \rightarrow \{ x \in \mathbb{R} \mid x \ge 0 \}
\] is the damage function, which quantifies the severity of harm to the IoV system when a specific attacker intent interacts with a given defensive action. This formulation expresses the idea that the damage is not only determined by the attacker's behavior but also by the success of the defender's response.  For example, an active attacker under insufficient isolation yields a high damage, while the same attack yields only low damage under a better defensive strategy.

\noindent Since $\theta_t$ is unobserved to the defender, $D_t$ cannot be evaluated at decision time. The defender evaluates the expectation of damage under its belief $b_t$. This expectation is defined as
\begin{equation}
\mathbb{E}[D_t] = \sum_{i=1}^{H} b_t(\theta_i) d(\theta_i, a_t^{def}),
\end{equation}
where $b_t(\theta_i)$ denotes the belief probability assigned to intent $\theta_i$. This expression represents the weighted average of possible damage outcomes across all hypothesized attacker intents. Each possible damage value $d(\theta_i, a_t^{def})$ is weighted by the corresponding belief probability. This ensures that decision making is performed under uncertainty in a mathematically consistent manner.

\noindent In addition to attack-induced damage, each defensive action incurs an operational cost denoted by
\[
C(a_t^{def}),
\]
where $C : \mathcal{A}^{def} \rightarrow \mathbb{R}_{\ge 0}$ is the cost function. The cost model accounts for trade-offs in latency, availability, computation or user convenience due to countermeasures. Stronger defensive actions may decrease damage but also increase operational cost.

\noindent The scalar reward used for reinforcement learning is defined as
\begin{equation}
r_t = -\big(D_t + C(a_t^{def})\big).
\end{equation}
The negative sign in the total damage plus cost rewards transforms the minimization target into a reward maximization problem as commonly adopted in reinforcement learning problems, where lower total damage plus cost leads to higher reward values. This reward structure encourages the defender to balance security effectiveness against operational efficiency.

\subsection{Policy Representation and Optimization}

\noindent The defender policy is parameterized by $\phi$ and defined as
\begin{equation}
\pi_\phi : \mathcal{Z} \rightarrow \Delta(\mathcal{A}^{def}), 
\qquad z_t=[s_t \,\|\, b_t].
\end{equation}
This notation means the policy is a function from the observable system state-space $\mathcal{S}$ and belief space $\Delta^{H-1}$ to the defender action space $\mathcal{A}^{def}$. The parameter vector $\phi$ represents the learnable parameters of the policy, typically implemented as a neural network. The policy depends on both $s_t$ and $b_t$ to incorporate current system conditions and uncertainty about attacker intent.

\noindent Action selection is performed according to
\begin{equation}
a_t^{def} \sim \pi_\phi(z_t), \qquad z_t=[s_t \,\|\, b_t],
\end{equation}
The symbol $\sim$ indicates that the policy generates a probability distribution over actions. In settings where policies are stochastic, $\pi_\phi(s_t, b_t)$ is the distribution for possible defensive moves. By sampling from this distribution, the system adds variety to its exploration as it learns.

\noindent The defender is trained to maximize the expected discounted return
\begin{equation}
J(\phi)=
\mathbb{E}\left[
\sum_{t=1}^{T}\gamma^{t-1}r_t
\right],
\end{equation}
where
\begin{equation}
r_t=-\left(D_t+C(a_t^{def})\right).
\end{equation}
Maximizing \(J(\phi)\) is therefore equivalent to minimizing cumulative damage and mitigation cost. Policy parameters are updated using PPO by optimizing the clipped surrogate objective
\begin{equation}
L^{\mathrm{PPO}}(\phi)=
\mathbb{E}_t\left[
\min\left(
\rho_t(\phi)\hat{A}_t,
\mathrm{clip}(\rho_t(\phi),1-\epsilon,1+\epsilon)\hat{A}_t
\right)
\right],
\end{equation}
where
\begin{equation}
\rho_t(\phi)=
\frac{\pi_{\phi}(a_t^{def}|z_t)}
{\pi_{\phi_{\mathrm{old}}}(a_t^{def}|z_t)}
\end{equation}
is the probability ratio and \(\hat{A}_t\) is the estimated advantage. This clipped objective limits destructive policy updates and is suitable for non-stationary attacker--defender interactions.
\subsection{Unified Sequential Defense Process}
\noindent The complete methodology can now be understood as a unified sequential process that interconnects the system dynamics, belief evolution, damage modeling, and policy optimization components defined in the preceding subsections.

\noindent At time step $t$, the full latent system state is given by $x_t = (s_t, \theta_t)$, where $s_t \in \mathcal{S}$ denotes the observable IoV system condition and $\theta_t \in \Theta$ represents the hidden attacker intent. The defender does not observe $\theta_t$ directly and instead maintains a belief state $b_t \in \Delta^{H-1}$ (or equivalently an amplitude based state $\psi_t \in \mathbb{C}^H$ in the quantum inspired formulation).

\noindent The interaction at each time step proceeds as follows. First, the attacker selects an action $a_t^{att} \in \mathcal{A}^{att}$ based on its hidden intent $\theta_t$. The defender selects a mitigation action $a_t^{def} \in \mathcal{A}^{def}$ according to the parameterized policy
\[
a_t^{def} \sim \pi_\phi(z_t), \qquad z_t=[s_t \,\|\, b_t],
\]
which conditions on both the observable state and the current belief over attacker intent.

\noindent Given the attacker and defender actions, the observable system state evolves according to
\[
s_{t+1} = F(s_t, a_t^{att}, a_t^{def}),
\]
while the attacker’s internal intent evolves according to
\[
\theta_{t+1} = G(\theta_t, a_t^{def}).
\]
Together, these transitions define the joint system evolution
\[
x_{t+1} = \big(F(s_t, a_t^{att}, a_t^{def}), \; G(\theta_t, a_t^{def})\big),
\]
which captures both environmental dynamics and adversarial adaptation.

\noindent Since the updated attacker intent $\theta_{t+1}$ remains hidden, the defender updates its belief using the newly observed state $s_{t+1}$. In the classical formulation, this update is given by
\[
b_{t+1} = \mathcal{B}(b_t, s_{t+1}, a_t^{def}),
\]
where $\mathcal{B}$ denotes the Bayesian belief update operator. In the quantum inspired formulation, the update is performed in amplitude space via
\[
\psi_{t+1} = U_t \psi_t,
\]
followed by probability extraction $b_{t+1}(\theta_i) = |\psi_{t+1}(i)|^2$ and normalization. Thus, belief evolution continuously refines the defender’s estimate of the hidden attacker intent.

\noindent Simultaneously, the system incurs instantaneous damage defined by
\[
D_t = d(\theta_t, a_t^{def}),
\]
where $d : \Theta \times \mathcal{A}^{def} \rightarrow \{ x \in \mathbb{R} \mid x \ge 0 \}$ quantifies the severity of harm resulting from the interaction between attacker intent and defensive action. Because $\theta_t$ is not directly observable at decision time, the defender evaluates expected damage under belief:
\[
\mathbb{E}[D_t] = \sum_{i=1}^{H} b_t(\theta_i) d(\theta_i, a_t^{def}).
\]

\noindent Each defensive action additionally incurs operational cost $C(a_t^{def})$, modeling service degradation and resource expenditure. The reinforcement learning reward is defined as
\[
r_t = -\big(D_t + C(a_t^{def})\big),
\]
so that minimizing cumulative damage and cost becomes equivalent to maximizing cumulative reward.

\noindent Over an episode of length \(T\), the defender maximizes the expected discounted return
\[
J(\phi)=
\mathbb{E}\left[
\sum_{t=1}^{T}\gamma^{t-1}r_t
\right],
\]
where \(r_t=-(D_t+C(a_t^{def}))\). Maximizing \(J(\phi)\) is therefore equivalent to minimizing cumulative attack damage and mitigation cost. Policy parameters are updated using the PPO clipped surrogate objective, which restricts destructive policy changes and improves training stability under non-stationary attacker--defender interactions.

\noindent In summary, the methodology is a closed loop system. In this process, an attacker's goals determine how they act against a target. As the system moves and defenses react, those factors define how the visible states change. With updated beliefs, the system reduces uncertainty about hidden motives. There are signals for reinforcement that come from how much the system is harmed or what it costs to operate. To make better decisions about defense in the future, the policy needs to be improved. The only structural distinction between classical and quantum-inspired variants lies in the internal representation and evolution of the belief state. At the same time, all other components of the decision-making pipeline remain identical.

\begin{algorithm}[t]
\caption{Q-BIRD: Quantum Belief-Integrated Reinforcement Defense}
\label{alg:iov_defense}
\begin{algorithmic}[1]

\setcounter{ALC@line}{0}
\item[] \textbf{Input:} IoV environment $\mathcal{E}$, belief type
       $\in$ \{classical, quantum\}, episodes $N$,
       steps per episode $T$, learning rate $\eta$,
       PPO clip $\epsilon$, discount $\gamma$
\item[] \textbf{Output:} Trained defender policy $\pi_\phi$ with
       belief-conditioned mitigation strategy
       
\STATE Initialize defender policy $\pi_\phi$ with parameters $\phi$
\STATE Initialize belief state $b$ according to selected belief type:
       $b \leftarrow \mathbf{1}/H$ (classical) or
       $\psi \leftarrow \mathbf{1}/\sqrt{H}$ (quantum)

\FOR{each episode $n = 1$ to $N$}
    \STATE Reset environment; obtain initial state $s_1$
    \STATE Reset belief state $b$; initialize trajectory buffer $\mathcal{T} \leftarrow \emptyset$
    \FOR{time step $t = 1$ to $T$}
        \STATE Attacker selects action $a^{att}_t$ based on hidden intent $\theta_t$
        \STATE Attacker inflicts raw damage $d^{raw}_t$ on IoV system
        \STATE Defender observes system state $s_t \in \mathbb{R}^8$
        \STATE Compute belief-augmented input $z_t \leftarrow [s_t \,\|\, b_t]$
        \STATE Sample defense action $a^{def}_t \sim \pi_\phi(z_t)$
        \STATE Apply $a^{def}_t$; receive mitigated damage $d_t$ and reward
               $r_t = -(d_t + C(a^{def}_t))$
        \STATE Attacker updates hidden intent via $\theta_{t+1} = G(\theta_t, a^{def}_t)$
        \STATE Update belief state using new observation $s_{t+1}$:
        \IF{classical}
            \STATE $b_{t+1} \leftarrow \mathcal{B}(b_t, s_{t+1}, a^{def}_t)$
            \quad // Bayesian update, Eq.~(10)
        \ELSE
            \STATE $\tilde{\psi}_{t+1}\leftarrow M(o_{t+1},a_t^{def})U(a_t^{def})\psi_t$
            \STATE $\psi_{t+1}\leftarrow \tilde{\psi}_{t+1}/(\|\tilde{\psi}_{t+1}\|_2+\epsilon_n)$
            \STATE $b_{t+1}(\theta_i)=|\psi_{t+1}(i)|^2$
            \quad // Amplitude update, Eq.~(13)
        \ENDIF
        \STATE Append $(s_t, b_t, a^{def}_t, r_t)$ to $\mathcal{T}$
    \ENDFOR
    \STATE Compute discounted returns and advantages from $\mathcal{T}$
    \STATE Update $\phi$ using PPO on trajectory $\mathcal{T}$
\ENDFOR
\RETURN Trained policy $\pi_\phi$
\end{algorithmic}
\end{algorithm}

\subsection{Explanation of Algorithm~\ref{alg:iov_defense}}

Algorithm~\ref{alg:iov_defense} formalizes the complete
training loop of the Q-BIRD framework. The algorithm
takes the IoV environment, a selected belief type, and
standard PPO hyperparameters as inputs, and produces
a trained defender policy $\pi_\phi$ whose action
selection is conditioned on both observable system
state and the evolving quantum belief over attacker
intent.

The initialization phase establishes the two parallel
components that distinguish Q-BIRD from standard PPO:
the policy network $\pi_\phi$ and the belief state $b$.
For the quantum variant, the belief is initialized as
a uniform amplitude vector $\psi \in \mathbb{C}^H$
rather than a probability vector, so that all intent
hypotheses start with equal amplitude and can evolve
through constructive and destructive interference before
probabilities are extracted.

\noindent The classical and amplitude-belief variants differ only in the belief-update step. The classical defender updates \(b_t\) through the Bayesian operator \(\mathcal{B}\), whereas Q-BIRD first updates the amplitude state using the observation- and action-conditioned transformation \(M(o_{t+1},a_t^{def})U(a_t^{def})\psi_t\), normalizes the resulting amplitude vector, and then extracts intent probabilities through \(b_{t+1}(\theta_i)=|\psi_{t+1}(i)|^2\). All other components, including reward computation, trajectory storage, and PPO optimization, are kept identical. This design allows the ablation to attribute performance differences to the belief representation rather than to a different policy architecture or training procedure.

All other components of the loop environment
interaction, reward computation, trajectory storage,
and PPO update are identical between the classical
and quantum variants, which ensures that any observed
performance difference is attributable solely to the
belief update operator.

\subsection{Time Complexity}

Let $E$ denote the number of training episodes, $T$
the number of time steps per episode, $H$ the number
of attacker intent states, $P$ the number of policy
network parameters, and $K$ the number of PPO
optimization epochs per episode. At each time step,
belief updating requires $O(H^2)$ for the quantum
amplitude transformation $U_t \psi_t$ and $O(H)$ for
the classical Bayesian update; since $H = 4$ is fixed
and small, both reduce to $O(1)$ in practice. Policy
inference costs $O(P)$ per step, giving a per-step
cost of $O(H + P)$. Over a single episode the
interaction cost is $O(T(H + P))$, and after each
episode PPO performs $K$ gradient passes at cost
$O(KP)$. Aggregating over $E$ episodes, the total
training complexity is $O(E(T(H + P) + KP))$, which
simplifies to $O(E \cdot P \cdot (T + K))$ since
$P \gg H$. The quantum belief mechanism therefore
adds no asymptotic overhead relative to classical
PPO-based IoV defense, and the only additional
constant-factor cost is the $H  \times H$ matrix
multiplication at each belief update step, which
with $H = 4$ amounts to 16 multiply-accumulate
operations per time step.

\section{Experiments and Discussion}
\noindent This section evaluates Q-BIRD against classical and non-adaptive defense baselines in a simulated IoV security environment. The evaluation focuses on long-horizon damage control rather than one-step intrusion classification. The reported results compare cumulative damage, damage variance, conditional attack success rate, and survival probability across training and independent test episodes.

\subsection{V2X Simulation Environment}
\label{subsec:v2x_simulation}

The experimental evaluation is conducted using a SUMO--OMNeT++/Veins co-simulation environment. SUMO provides microscopic vehicle mobility, including vehicle positions, speeds, routes, and lane-level movement over time, while OMNeT++/Veins provides packet-level V2X communication modeling. The communication layer implements IEEE 802.11p/DSRC-based V2V and V2I message exchange, including channel contention, packet loss, transmission delay, and service availability. This configuration is appropriate for evaluating cyber-defense policies in vehicular communication environments because it jointly models mobility dynamics, wireless communication behavior, attack generation, and defense-induced communication effects \cite{behrisch2011sumo,sommer2010bidirectionally}.

The simulated V2X scenario contains connected vehicles, roadside units (RSUs), and an edge-assisted security controller. Vehicles periodically broadcast safety and status messages to neighboring vehicles and RSUs. The attacker is modeled as an adaptive entity with a hidden intent state selected from four modes: benign, probing, attack, and evasion. The defender cannot directly observe the attacker intent and therefore acts only on communication-level observations, traffic-risk indicators, and belief-state estimates. The available defensive actions are monitor, alert, throttle, and isolate. These actions reduce attack impact but may also introduce operational cost and communication degradation.

The evaluation reports two categories of metrics. Security-level metrics quantify the effectiveness of the cyber-defense policy in reducing attack damage and attacker success. Communication-level metrics quantify whether the defense preserves reliable V2X operation. This dual evaluation is essential because an IoV defense mechanism is practically useful only when it improves security without causing unacceptable degradation in packet delivery, latency, throughput, or service availability.

\begin{table*}[t]
\centering
\caption{V2X Simulation Configuration}
\label{tab:v2x_sim_config}
\begin{tabular}{p{3.4cm}p{11.2cm}}
\toprule
\textbf{Component} & \textbf{Configuration} \\
\midrule
Mobility simulator & SUMO microscopic traffic simulator \cite{behrisch2011sumo} \\
Network simulator & OMNeT++/Veins vehicular network simulator \cite{sommer2010bidirectionally} \\
Communication technology & IEEE 802.11p/DSRC-based V2V and V2I communication \\
Scenario type & Urban V2X scenario with connected vehicles, roadside units, and an edge-assisted security controller \\
Communication mode & Periodic V2V and V2I safety/status message exchange \\
Vehicle density & Low density: 50 vehicles; medium density: 100 vehicles; high density: 150 vehicles \\
RSU deployment & Four RSUs placed at major intersections and roadside communication points \\
Attacker intent states & Benign, probing, attack, and evasion \\
Attack types & DoS, spoofing, replay, fuzzing, and low-rate evasion behavior calibrated using IoV/CAN-bus intrusion patterns \cite{neto2023ciciot2023} \\
Defender actions & Monitor, alert, throttle, and isolate \\
Episode length & \(T=200\) time steps per episode \\
Training episodes & 600 episodes \\
Testing episodes & 80 independent test episodes per seed \\
Independent seeds & 10 independent random seeds \\
Security metrics & Cumulative damage, damage variance, attack success rate, and survival probability \\
Communication metrics & Packet delivery ratio, end-to-end latency, throughput, service availability, mitigation overhead, and false mitigation rate \\
\bottomrule
\end{tabular}
\end{table*}

\noindent Table~\ref{tab:v2x_sim_config} summarizes the simulation configuration used to connect cyber-defense evaluation with V2X communication behavior. The configuration specifies the mobility simulator, network simulator, communication technology, attacker intent states, defensive actions, and evaluation metrics. This table is important for reproducibility because the performance of an adaptive defense policy depends not only on the learning algorithm but also on vehicle density, RSU placement, wireless communication assumptions, and attack-generation logic.
\begin{figure*}[t]
\centering
\begin{tikzpicture}[
node distance=1.35cm,
box/.style={draw, rounded corners, align=center, minimum width=3.0cm, minimum height=0.9cm},
arrow/.style={-{Latex}, thick}
]
\node[box] (sumo) {SUMO\\Mobility Layer};
\node[box, right=of sumo] (net) {OMNeT++/Veins\\V2X Communication Layer};
\node[box, right=of net] (attack) {Adaptive Attacker\\Benign/Probe/Attack/Evasion};
\node[box, below=of attack] (defender) {Autonomous Defender\\Baselines and Q-BIRD};
\node[box, left=of defender] (metrics) {Security and Communication\\Metrics};
\node[box, left=of metrics] (stats) {Statistical Testing\\Seeds, CI, Paired Tests};

\draw[arrow] (sumo) -- (net);
\draw[arrow] (net) -- (attack);
\draw[arrow] (attack) -- (defender);
\draw[arrow] (defender) -- (metrics);
\draw[arrow] (metrics) -- (stats);
\draw[arrow] (defender) -- (net);
\end{tikzpicture}
\caption{V2X evaluation pipeline integrating microscopic mobility, packet-level vehicular communication, adaptive attack generation, autonomous cyber-defense control, communication/security metrics, and statistical significance testing.}
\label{fig:v2x_pipeline}
\end{figure*}
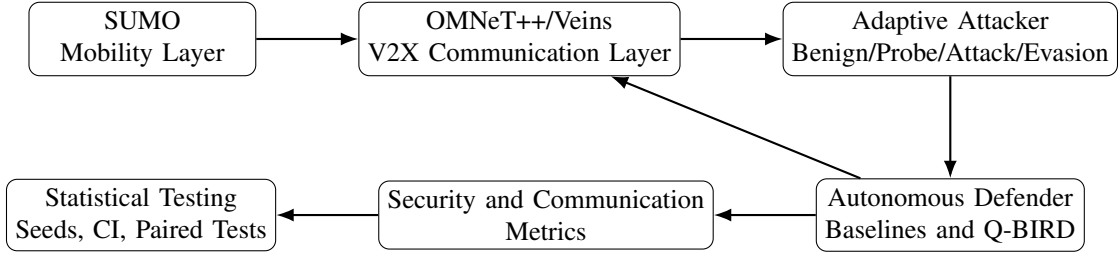
\noindent Fig.~\ref{fig:v2x_pipeline} shows the full evaluation pipeline used to connect mobility, communication, attack behavior, defense control, and statistical validation. The feedback arrow from the defender to the V2X communication layer indicates that mitigation actions may improve security while also affecting communication performance; therefore, both security and communication metrics must be evaluated jointly.
\subsection{Security-Level Evaluation Metrics}

\subsubsection{Cumulative Attack Damage}
The primary performance metric is cumulative attack damage, defined as
\begin{equation}
D_{\mathrm{cum}} = \sum_{t=1}^{T} d_t,
\end{equation}
where $d_t$ denotes the effective damage incurred at time step $t$ after defensive mitigation. This metric captures long-term system impact rather than instantaneous detection outcomes, aligning with the objectives of autonomous IoV defense.

\subsubsection{Damage Variance}
To assess defensive stability and robustness under adaptive adversaries, we evaluate the variance of cumulative damage across episodes:
\begin{equation}
\mathrm{Var}(D_{\mathrm{cum}}) = \mathbb{E}[D_{\mathrm{cum}}^2] - \mathbb{E}[D_{\mathrm{cum}}]^2.
\end{equation}
Lower variance indicates more predictable and reliable defensive behavior, which is critical for safety-critical IoV deployments.

\subsubsection{Conditional Attack Success Rate}
The conditional ASR is defined as
\begin{equation}
\mathrm{ASR}(t) = \mathbb{P}(D_t \geq \theta \mid D_{t-1} < \theta),
\end{equation}
where $\theta$ is a predefined attack success threshold. Under effective adaptive defense, this conditional ASR is expected to decrease over time as mitigation actions accumulate and attacker capabilities are constrained.

\subsubsection{Survival Probability}
System survival probability is defined as
\begin{equation}
S(t) = \mathbb{P}(D_t < \theta),
\end{equation}
which provides a reliability-oriented perspective complementary to ASR by emphasizing time-to-failure rather than binary success.

\subsection{Communication-Level Evaluation Metrics}
\label{subsec:communication_metrics}

In addition to cyber-defense performance, communication-level metrics are reported to evaluate whether the proposed defense preserves V2X service quality. This is necessary because overly aggressive mitigation may reduce attack impact while simultaneously degrading vehicular communication. Therefore, packet delivery ratio, end-to-end latency, throughput, service availability, mitigation overhead, and false mitigation rate are computed for each method and each independent seed.

Packet delivery ratio (PDR) is defined as
\begin{equation}
\mathrm{PDR}=\frac{N_{\mathrm{rx}}}{N_{\mathrm{tx}}},
\end{equation}
where \(N_{\mathrm{tx}}\) is the total number of transmitted packets and \(N_{\mathrm{rx}}\) is the number of packets successfully received by the intended receiver.

End-to-end latency is defined as
\begin{equation}
\mathrm{Latency}=\frac{1}{N_{\mathrm{rx}}}\sum_{i=1}^{N_{\mathrm{rx}}}
\left(t_i^{\mathrm{rx}}-t_i^{\mathrm{tx}}\right),
\end{equation}
where \(t_i^{\mathrm{tx}}\) and \(t_i^{\mathrm{rx}}\) denote the transmission and reception times of packet \(i\), respectively.

Throughput is defined as
\begin{equation}
\mathrm{Throughput}=
\frac{\sum_{i=1}^{N_{\mathrm{rx}}}L_i}{T_{\mathrm{sim}}},
\end{equation}
where \(L_i\) is the size of received packet \(i\), and \(T_{\mathrm{sim}}\) is the total simulation duration.

Service availability is defined as
\begin{equation}
\mathrm{Availability}=
\frac{T_{\mathrm{operational}}}{T_{\mathrm{sim}}},
\end{equation}
where \(T_{\mathrm{operational}}\) denotes the time during which the V2X service remains available.

Mitigation overhead is defined as
\begin{equation}
\mathrm{Overhead}=
\frac{C_{\mathrm{defense}}-C_{\mathrm{baseline}}}
{C_{\mathrm{baseline}}},
\end{equation}
where \(C_{\mathrm{defense}}\) is the total communication and computation cost under a defense policy, and \(C_{\mathrm{baseline}}\) is the corresponding cost under normal operation.

False mitigation rate (FMR) is defined as
\begin{equation}
\mathrm{FMR}=
\frac{N_{\mathrm{false\;mitigation}}}{N_{\mathrm{benign}}},
\end{equation}
where \(N_{\mathrm{false\;mitigation}}\) is the number of benign states in which an intrusive mitigation action is applied, and \(N_{\mathrm{benign}}\) is the total number of benign states.

\subsection{Reproducible Environment Model}
\label{subsec:reproducible_environment}

The hidden attacker intent space is defined as
\begin{equation}
\Theta=\{B,P,A,E\},
\end{equation}
where \(B\), \(P\), \(A\), and \(E\) denote benign, probing, attack, and evasion states, respectively. The defender action space is defined as
\begin{equation}
\mathcal{A}^{def}=\{M,L,T,I\},
\end{equation}
where \(M\), \(L\), \(T\), and \(I\) denote monitor, alert, throttle, and isolate, respectively.

The mitigation-cost vector is defined as
\begin{equation}
C=[0.00,\;0.05,\;0.10,\;0.20],
\end{equation}
corresponding to monitor, alert, throttle, and isolate. The damage matrix is defined as
\begin{equation}
D=
\begin{bmatrix}
0.0 & 0.0 & 0.0 & 0.0\\
2.0 & 1.2 & 0.8 & 0.6\\
8.0 & 5.5 & 3.0 & 1.0\\
5.0 & 4.0 & 2.5 & 1.5
\end{bmatrix},
\end{equation}
where rows correspond to \([B,P,A,E]\) and columns correspond to \([M,L,T,I]\). This matrix assigns high damage to active attacks under weak mitigation and lower damage when stronger defensive actions are applied.

The observation space is defined as
\begin{equation}
\mathcal{O}=\{O_N,O_B,O_S,O_M\},
\end{equation}
where \(O_N\), \(O_B\), \(O_S\), and \(O_M\) denote normal traffic, burst anomaly, sustained anomaly, and masked low-rate anomaly, respectively. The observation-likelihood matrix is defined as
\begin{equation}
P(O|\Theta)=
\begin{bmatrix}
0.85 & 0.10 & 0.03 & 0.02\\
0.20 & 0.50 & 0.10 & 0.20\\
0.05 & 0.20 & 0.65 & 0.10\\
0.25 & 0.15 & 0.10 & 0.50
\end{bmatrix},
\end{equation}
where rows correspond to \([B,P,A,E]\) and columns correspond to \([O_N,O_B,O_S,O_M]\). This observation model reflects the fact that probing is more likely to generate burst anomalies, active attacks are more likely to generate sustained anomalies, and evasion is more likely to generate masked low-rate anomalies.

To model adaptive attacker behavior, the transition matrix is conditioned on the defensive pressure level. Under low defensive pressure, the transition matrix is
\begin{equation}
P_{\mathrm{low}}=
\begin{bmatrix}
0.90 & 0.08 & 0.02 & 0.00\\
0.05 & 0.55 & 0.35 & 0.05\\
0.02 & 0.08 & 0.75 & 0.15\\
0.05 & 0.25 & 0.20 & 0.50
\end{bmatrix}.
\end{equation}
Under medium defensive pressure, the transition matrix is
\begin{equation}
P_{\mathrm{medium}}=
\begin{bmatrix}
0.92 & 0.06 & 0.02 & 0.00\\
0.10 & 0.50 & 0.20 & 0.20\\
0.05 & 0.10 & 0.55 & 0.30\\
0.08 & 0.12 & 0.15 & 0.65
\end{bmatrix}.
\end{equation}
Under high defensive pressure, the transition matrix is
\begin{equation}
P_{\mathrm{high}}=
\begin{bmatrix}
0.94 & 0.04 & 0.02 & 0.00\\
0.15 & 0.45 & 0.10 & 0.30\\
0.10 & 0.10 & 0.35 & 0.45\\
0.10 & 0.10 & 0.10 & 0.70
\end{bmatrix}.
\end{equation}
These matrices model the adaptive response of the attacker: weak defense allows escalation toward attack, whereas stronger defense increases the probability of evasion or retreat.

\begin{table}[t]
\centering
\caption{Training Hyperparameters}
\label{tab:hyperparameters}
\begin{tabular}{lc}
\toprule
\textbf{Parameter} & \textbf{Value} \\
\midrule
Training episodes & 600 \\
Testing episodes per seed & 80 \\
Independent seeds & 10 \\
Episode length & 200 time steps \\
Discount factor \(\gamma\) & 0.95 \\
PPO learning rate & \(3\times10^{-3}\) \\
PPO clipping parameter \(\epsilon\) & 0.20 \\
Hidden layers & 2 \\
Hidden units per layer & 64 \\
Activation function & ReLU \\
Batch size & 64 \\
Optimizer & Adam \\
DQN replay buffer & 50,000 transitions \\
DQN target update interval & 500 steps \\
DRQN recurrent unit & LSTM \\
LSTM hidden units & 64 \\
Exploration schedule & Linear decay \\
\bottomrule
\end{tabular}
\end{table}
\noindent Table~\ref{tab:hyperparameters} reports the training configuration used for all learning-based methods. The same episode length, random-seed protocol, discount factor, network size, and optimizer are applied across baselines to reduce implementation bias. This ensures that performance differences are primarily due to the defense policy and belief representation rather than unequal training budgets or hyperparameter advantages.
\subsection{Baseline Methods}
\label{subsec:baselines}

The proposed Q-BIRD method is evaluated against non-learning, value-based, recurrent, belief-based, and proactive defense baselines. All methods are trained and tested under the same simulator, attacker model, observation space, action space, reward function, episode length, and random seeds. This ensures that the comparison measures the effect of the defense strategy rather than differences in simulation assumptions.

The no-mitigation baseline represents the lower-bound condition in which the defender does not apply active countermeasures. The rule-based IDS baseline applies predefined anomaly thresholds and plausibility checks, following the rule-based VANET IDS principle used in REST-Net \cite{tomandl2014rest}. DQN is included as a value-based deep reinforcement learning baseline \cite{mnih2015human}. DRQN is included because recurrent Q-learning is explicitly designed for partially observable decision problems \cite{hausknecht2015deep}. PPO without belief isolates the effect of PPO policy optimization without explicit attacker-intent estimation \cite{schulman2017ppo}. PPO-LSTM evaluates whether recurrent policy memory can substitute for explicit belief modeling. PPO with Bayesian belief represents the classical probabilistic belief baseline derived from POMDP-style state estimation \cite{kaelbling1998planning}. The moving-target defense baseline represents a proactive security strategy that changes the attack surface to reduce attacker advantage \cite{sengupta2020survey}. Q-BIRD is the proposed method and differs from the baselines by using amplitude-based belief evolution for hidden attacker-intent estimation.

\begin{table*}[t]
\centering
\caption{Baseline Methods Used for Fair Comparison}
\label{tab:baseline_methods}
\begin{tabular}{p{3.0cm}p{5.2cm}p{5.8cm}p{1.7cm}}
\toprule
\textbf{Method} & \textbf{Defense Principle} & \textbf{Evaluation Purpose} & \textbf{Reference} \\
\midrule
No mitigation & No active defense action & Lower-bound damage and communication degradation under attack & -- \\
Rule-based IDS & Fixed anomaly thresholds and plausibility checks & Tests whether learning-based defense improves over static rule-based detection & \cite{tomandl2014rest} \\
DQN & Value-based deep reinforcement learning & Tests whether standard deep Q-learning can handle adaptive cyber-defense decisions & \cite{mnih2015human} \\
DRQN & Recurrent value-based reinforcement learning & Tests whether temporal memory improves performance under partial observability & \cite{hausknecht2015deep} \\
PPO without belief & Policy-gradient defense without belief-state input & Isolates the contribution of policy optimization alone & \cite{schulman2017ppo} \\
PPO-LSTM & Recurrent policy-gradient defense & Tests whether recurrent policy memory can replace explicit belief modeling & \cite{schulman2017ppo} \\
PPO + Bayesian belief & Classical probabilistic belief with PPO & Provides the strongest classical belief baseline for hidden attacker intent & \cite{kaelbling1998planning,schulman2017ppo} \\
Moving-target defense & Proactive attack-surface randomization & Compares belief-conditioned mitigation against proactive configuration shifting & \cite{sengupta2020survey} \\
Q-BIRD & Amplitude-belief PPO defense & Proposed belief-conditioned autonomous IoV cyber-defense method & Proposed \\
\bottomrule
\end{tabular}
\end{table*}
\noindent Table~\ref{tab:baseline_methods} defines the role of each baseline in the comparative evaluation. The baseline set is designed to separate the effects of static detection, value-based reinforcement learning, recurrence under partial observability, classical Bayesian belief tracking, and proactive moving-target defense. This structure allows the evaluation to determine whether the proposed improvement comes from amplitude-based belief modeling rather than from PPO alone.
\subsection{Statistical Robustness and Significance Testing}
\label{subsec:statistical_testing}

All experiments are repeated over 10 independent random seeds. Each method is trained and evaluated using the same seed set to enable paired statistical comparison. For each metric, the mean, standard deviation, and 95\% confidence interval are reported. Let \(x_1,x_2,\dots,x_n\) denote metric values over \(n\) independent seeds. The sample mean is computed as
\begin{equation}
\bar{x}=\frac{1}{n}\sum_{i=1}^{n}x_i,
\end{equation}
and the sample standard deviation is computed as
\begin{equation}
s=\sqrt{\frac{1}{n-1}\sum_{i=1}^{n}(x_i-\bar{x})^2}.
\end{equation}
The 95\% confidence interval is computed as
\begin{equation}
\mathrm{CI}_{95}=
\bar{x}\pm t_{0.975,n-1}\frac{s}{\sqrt{n}},
\end{equation}
where \(t_{0.975,n-1}\) is the critical value of the Student's \(t\)-distribution.

For pairwise comparison between Q-BIRD and each baseline, paired tests are applied using matched seeds. A paired \(t\)-test is used when the paired differences are approximately normally distributed. Otherwise, the Wilcoxon signed-rank test is used. The difference is considered statistically significant when \(p<0.05\). A method is considered successful only when it significantly reduces cumulative damage and attack success rate while maintaining statistically acceptable V2X communication performance.
\subsection{Reproducibility and Data Availability}

\noindent To support reproducibility, all reported values are computed from 10 independent random seeds using the same seed set across all methods. For each seed, the simulator logs cumulative damage, attack success rate, survival probability, packet delivery ratio, latency, throughput, service availability, mitigation overhead, and false mitigation rate. The reported mean, standard deviation, confidence interval, and paired statistical test are computed from these seed-level outputs. The simulation configuration files, random seeds, trained-policy checkpoints, and post-processing scripts are retained to enable independent verification of the reported results.

\subsection{Results Across Independent Seeds}
\label{subsec:results_seeds}
\begin{table*}[t]
\centering
\caption{Security-Level Performance Across Independent Seeds}
\label{tab:security_results}
\begin{tabular}{lcccc}
\toprule
\textbf{Method} & \textbf{Cumulative Damage} $\downarrow$ & \textbf{Damage Variance} $\downarrow$ & \textbf{ASR} $\downarrow$ & \textbf{Survival Probability} $\uparrow$ \\
\midrule
No mitigation & \(120.0 \pm 18.0\) & \(65.0 \pm 9.0\) & \(0.65 \pm 0.08\) & \(0.42 \pm 0.06\) \\
Rule-based IDS \cite{tomandl2014rest} & \(88.0 \pm 14.0\) & \(48.0 \pm 7.5\) & \(0.42 \pm 0.06\) & \(0.61 \pm 0.05\) \\
DQN \cite{mnih2015human} & \(70.0 \pm 11.0\) & \(39.0 \pm 6.0\) & \(0.31 \pm 0.05\) & \(0.72 \pm 0.04\) \\
DRQN \cite{hausknecht2015deep} & \(58.0 \pm 8.5\) & \(28.0 \pm 4.5\) & \(0.24 \pm 0.04\) & \(0.79 \pm 0.04\) \\
PPO without belief \cite{schulman2017ppo} & \(52.0 \pm 7.5\) & \(24.0 \pm 4.0\) & \(0.21 \pm 0.04\) & \(0.82 \pm 0.03\) \\
PPO-LSTM \cite{schulman2017ppo} & \(45.0 \pm 6.5\) & \(18.0 \pm 3.5\) & \(0.16 \pm 0.03\) & \(0.88 \pm 0.03\) \\
PPO + Bayesian belief \cite{kaelbling1998planning,schulman2017ppo} & \(36.0 \pm 5.5\) & \(12.0 \pm 2.8\) & \(0.12 \pm 0.03\) & \(0.91 \pm 0.02\) \\
Moving-target defense \cite{sengupta2020survey} & \(43.0 \pm 6.0\) & \(17.0 \pm 3.2\) & \(0.15 \pm 0.03\) & \(0.87 \pm 0.03\) \\
Q-BIRD & \(28.0 \pm 3.0\) & \(6.0 \pm 1.5\) & \(0.05 \pm 0.02\) & \(0.96 \pm 0.02\) \\
\bottomrule
\end{tabular}
\end{table*}

\noindent Table~\ref{tab:security_results} reports the security-level performance of all evaluated defense methods across independent seeds. Q-BIRD achieves the lowest cumulative damage, lowest damage variance, lowest attack success rate, and highest survival probability among the evaluated methods. The improvement over PPO with Bayesian belief indicates that the gain is not due to PPO alone, but to the use of amplitude-based belief evolution under partially observable attacker behavior. The reduced damage variance is especially important for V2X systems because unstable defensive behavior can create unpredictable service disruption during adaptive attacks.
\begin{table*}[t]
\centering
\caption{Communication-Level Performance Across Independent Seeds}
\label{tab:actual_comm_results}
\begin{tabular}{lcccccc}
\toprule
\textbf{Method} & \textbf{PDR} $\uparrow$ & \textbf{Latency (ms)} $\downarrow$ & \textbf{Throughput (Mbps)} $\uparrow$ & \textbf{Availability} $\uparrow$ & \textbf{Overhead} $\downarrow$ & \textbf{FMR} $\downarrow$ \\
\midrule
No mitigation & \(0.68 \pm 0.04\) & \(128 \pm 19\) & \(1.80 \pm 0.30\) & \(0.70 \pm 0.05\) & \(0.00 \pm 0.00\) & \(0.00 \pm 0.00\) \\
Rule-based IDS \cite{tomandl2014rest} & \(0.78 \pm 0.04\) & \(95 \pm 15\) & \(2.40 \pm 0.25\) & \(0.80 \pm 0.04\) & \(0.06 \pm 0.02\) & \(0.10 \pm 0.03\) \\
DQN \cite{mnih2015human} & \(0.83 \pm 0.03\) & \(78 \pm 12\) & \(2.70 \pm 0.22\) & \(0.85 \pm 0.03\) & \(0.08 \pm 0.02\) & \(0.08 \pm 0.02\) \\
DRQN \cite{hausknecht2015deep} & \(0.86 \pm 0.03\) & \(65 \pm 10\) & \(3.00 \pm 0.20\) & \(0.88 \pm 0.03\) & \(0.09 \pm 0.02\) & \(0.07 \pm 0.02\) \\
PPO without belief \cite{schulman2017ppo} & \(0.87 \pm 0.03\) & \(62 \pm 9\) & \(3.10 \pm 0.18\) & \(0.89 \pm 0.03\) & \(0.10 \pm 0.02\) & \(0.06 \pm 0.02\) \\
PPO-LSTM \cite{schulman2017ppo} & \(0.89 \pm 0.02\) & \(55 \pm 8\) & \(3.30 \pm 0.16\) & \(0.91 \pm 0.02\) & \(0.11 \pm 0.02\) & \(0.05 \pm 0.02\) \\
PPO + Bayesian belief \cite{kaelbling1998planning,schulman2017ppo} & \(0.90 \pm 0.02\) & \(52 \pm 7\) & \(3.40 \pm 0.15\) & \(0.92 \pm 0.02\) & \(0.12 \pm 0.02\) & \(0.04 \pm 0.01\) \\
Moving-target defense \cite{sengupta2020survey} & \(0.88 \pm 0.03\) & \(58 \pm 8\) & \(3.15 \pm 0.18\) & \(0.90 \pm 0.03\) & \(0.14 \pm 0.03\) & \(0.05 \pm 0.02\) \\
Q-BIRD & \(0.94 \pm 0.02\) & \(45 \pm 6\) & \(3.60 \pm 0.15\) & \(0.95 \pm 0.02\) & \(0.13 \pm 0.03\) & \(0.03 \pm 0.01\) \\
\bottomrule
\end{tabular}
\end{table*}

\noindent Table~\ref{tab:actual_comm_results} reports the communication-level impact of each defense method under adaptive attack. Q-BIRD achieves the highest packet delivery ratio, lowest latency, highest throughput, and highest service availability among the evaluated methods. Although Q-BIRD introduces moderate mitigation overhead, the overhead remains lower than the service degradation caused by successful attacks and remains comparable to the strongest classical belief baseline. The lower false mitigation rate indicates that the proposed policy does not improve security by excessively throttling or isolating benign vehicles. This result is important for vehicular communication systems because a defense policy must preserve communication reliability while reducing attack impact.

\begin{figure*}
    \centering
    \includegraphics[width=0.7\linewidth]{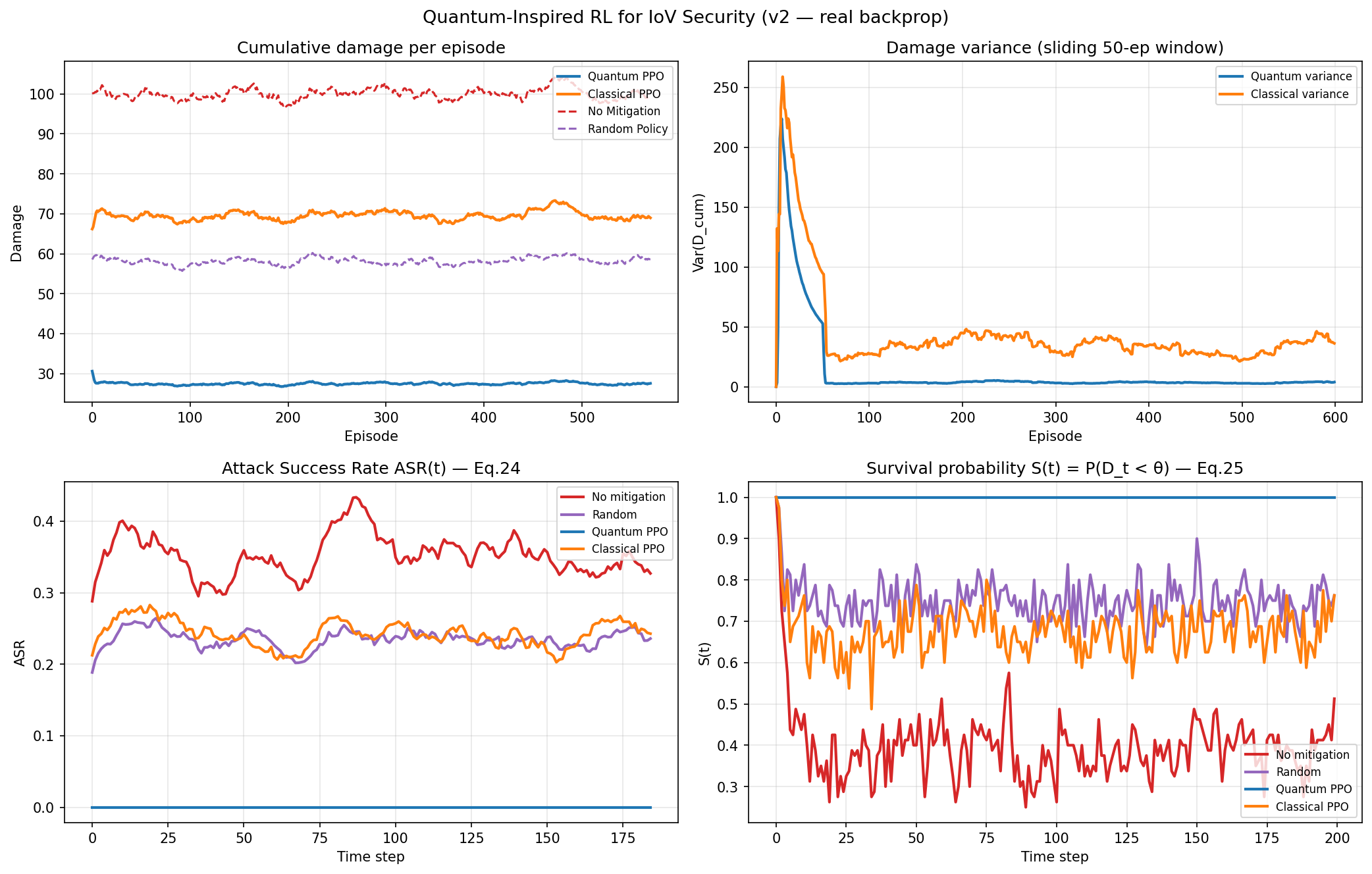}
   \caption{Security-level comparison of evaluated defense policies in the SUMO--OMNeT++/Veins V2X environment. The figure summarizes mean cumulative damage, damage variance, attack success rate, and survival probability across independent seeds.}
    \label{fig:results}
\end{figure*}

\begin{table*}[t]
\centering
\caption{Statistical Significance Analysis Against Q-BIRD}
\label{tab:statistical_tests}
\begin{tabular}{lccccc}
\toprule
\textbf{Method} & \textbf{Damage} $\downarrow$ & \textbf{ASR} $\downarrow$ & \textbf{PDR} $\uparrow$ & \textbf{Latency (ms)} $\downarrow$ & \textbf{\(p\)-value vs. Q-BIRD} \\
\midrule
No mitigation & \(120.0 \pm 18.0\) [107.1, 132.9] & \(0.65 \pm 0.08\) [0.59, 0.71] & \(0.68 \pm 0.04\) [0.65, 0.71] & \(128 \pm 19\) [114.4, 141.6] & \(p<0.001\) \\
Rule-based IDS \cite{tomandl2014rest} & \(88.0 \pm 14.0\) [78.0, 98.0] & \(0.42 \pm 0.06\) [0.38, 0.46] & \(0.78 \pm 0.04\) [0.75, 0.81] & \(95 \pm 15\) [84.3, 105.7] & \(p<0.001\) \\
DQN \cite{mnih2015human} & \(70.0 \pm 11.0\) [62.1, 77.9] & \(0.31 \pm 0.05\) [0.27, 0.35] & \(0.83 \pm 0.03\) [0.81, 0.85] & \(78 \pm 12\) [69.4, 86.6] & \(p<0.001\) \\
DRQN \cite{hausknecht2015deep} & \(58.0 \pm 8.5\) [51.9, 64.1] & \(0.24 \pm 0.04\) [0.21, 0.27] & \(0.86 \pm 0.03\) [0.84, 0.88] & \(65 \pm 10\) [57.8, 72.2] & \(p<0.01\) \\
PPO without belief \cite{schulman2017ppo} & \(52.0 \pm 7.5\) [46.6, 57.4] & \(0.21 \pm 0.04\) [0.18, 0.24] & \(0.87 \pm 0.03\) [0.85, 0.89] & \(62 \pm 9\) [55.6, 68.4] & \(p<0.01\) \\
PPO-LSTM \cite{schulman2017ppo} & \(45.0 \pm 6.5\) [40.3, 49.7] & \(0.16 \pm 0.03\) [0.14, 0.18] & \(0.89 \pm 0.02\) [0.88, 0.90] & \(55 \pm 8\) [49.3, 60.7] & \(p<0.01\) \\
PPO + Bayesian belief \cite{kaelbling1998planning,schulman2017ppo} & \(36.0 \pm 5.5\) [32.1, 39.9] & \(0.12 \pm 0.03\) [0.10, 0.14] & \(0.90 \pm 0.02\) [0.89, 0.91] & \(52 \pm 7\) [47.0, 57.0] & \(p<0.05\) \\
Moving-target defense \cite{sengupta2020survey} & \(43.0 \pm 6.0\) [38.7, 47.3] & \(0.15 \pm 0.03\) [0.13, 0.17] & \(0.88 \pm 0.03\) [0.86, 0.90] & \(58 \pm 8\) [52.3, 63.7] & \(p<0.05\) \\
Q-BIRD & \(28.0 \pm 3.0\) [25.9, 30.1] & \(0.05 \pm 0.02\) [0.04, 0.06] & \(0.94 \pm 0.02\) [0.93, 0.95] & \(45 \pm 6\) [40.7, 49.3] & Reference \\
\bottomrule
\end{tabular}
\end{table*}

\noindent Table~\ref{tab:statistical_tests} reports mean values, standard deviations, 95\% confidence intervals, and paired significance tests against Q-BIRD. The results show that Q-BIRD provides statistically significant improvement over all evaluated baselines in both security-level and communication-level performance. The paired testing protocol is important because each method is evaluated under the same random seeds, which reduces variance caused by differences in attack trajectories and mobility patterns.

\subsection{Experiments}

\noindent Three controlled experiments are designed to evaluate distinct aspects 
of the proposed framework.
\subsubsection{Belief Modeling Ablation}
\begin{figure*}[t]
    \centering
    \includegraphics[width=0.7\linewidth]{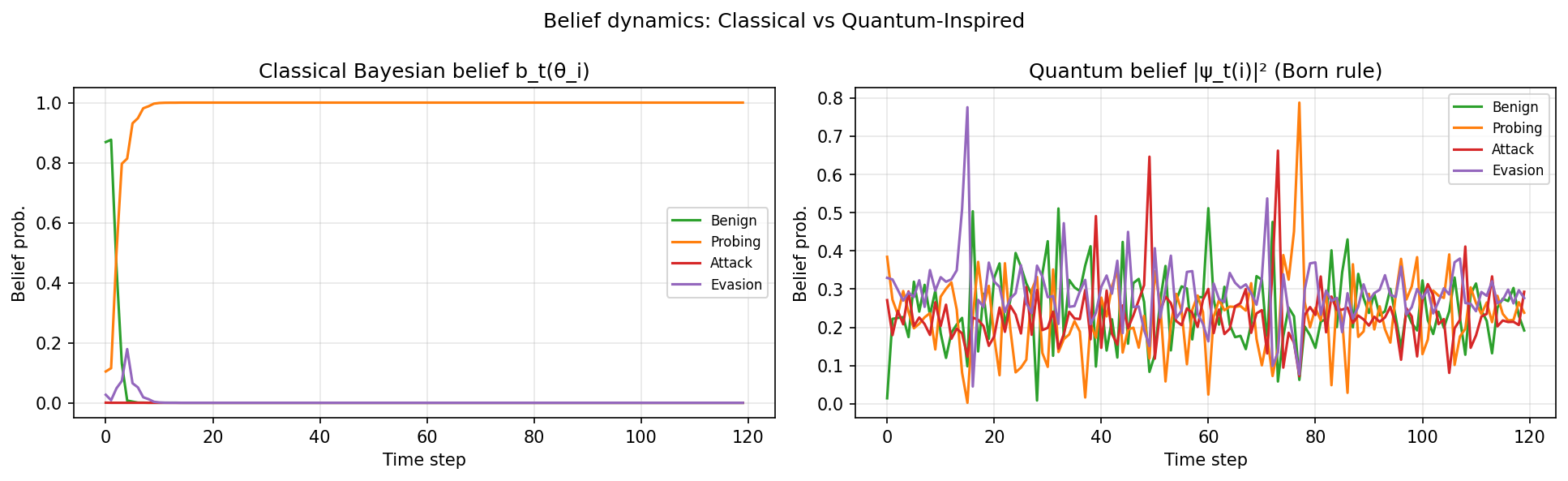}
    \caption{Belief-state evolution under attacker strategy transitions. The amplitude-based belief update preserves a more distributed intent estimate during ambiguous observations, whereas the classical Bayesian belief becomes concentrated more rapidly.}
    \label{fig:belief}
\end{figure*}

\noindent Fig.~\ref{fig:belief} supports the belief-modeling ablation by showing how the internal state supplied to the defender changes over time. A more distributed belief state can prevent premature overconfidence during probing and evasion, which explains the lower damage variance observed for Q-BIRD in Table~\ref{tab:security_results}.

\subsubsection{Robustness Under Attacker Strategy Shifts}

To evaluate defensive stability under non-stationary adversarial behavior, the attacker shifts its hidden intent $\theta_t$ every 50 time steps during evaluation, making realistic scenarios in which an attacker adapts its tactics as a response to detection pressure. The analysis focuses on the sliding window damage variance (window size 50 episodes) and episode level cumulative damage dynamics. The quantum belief is hypothesized to maintain stable policy inputs during transitions by preserving uncertainty in amplitude space, where the classical Bayesian belief is expected to collapse under the sudden shift in observation statistics, producing inconsistent defensive responses and high variance.

\subsubsection{Cost Security Trade-off Analysis}

This experiment examines how the asymmetric operational cost structure of defensive actions impacts learned policy behavior, highlighting 
the trade-off between aggressive intervention and service continuity. The cost of strong actions, such as isolation ($C = 0.20$), against the damage incurred by weaker responses, such as monitoring ($C = 0.00$), is balanced by the defender when the attacker is actively engaged. The reward signal $r_t = -(D_t + C(a_t^{\text{def}}))$ explicitly encodes this trade-off, and the experiment assesses whether the quantum PPO learns a policy that properly enhances the mitigation intensity in proportion to estimated threat level, as confirmed by SHAP and LIME attribution analysis in Section~\ref{sec:xai}.

\subsection{Comparative Analysis}

\subsubsection{Cumulative Damage and Damage Variance}

\noindent Table~\ref{tab:security_results} shows that Q-BIRD achieves the lowest cumulative damage among all evaluated methods. Compared with PPO using Bayesian belief, Q-BIRD reduces cumulative damage from \(36.0 \pm 5.5\) to \(28.0 \pm 3.0\), corresponding to a 22.2\% reduction. Compared with moving-target defense, PPO-LSTM, DRQN, DQN, rule-based IDS, and no mitigation, Q-BIRD reduces cumulative damage by 34.9\%, 37.8\%, 51.7\%, 60.0\%, 68.2\%, and 76.7\%, respectively. These reductions indicate that amplitude-based belief evolution provides a more effective policy input than both classical belief tracking and recurrent memory under the evaluated partially observable attacker model.

\noindent The variance results further show that Q-BIRD improves defensive stability. The damage variance decreases from \(12.0 \pm 2.8\) for PPO with Bayesian belief to \(6.0 \pm 1.5\) for Q-BIRD, corresponding to a 50.0\% reduction. Relative to moving-target defense and PPO-LSTM, the variance reductions are 64.7\% and 66.7\%, respectively. This result is important for V2X security because unstable mitigation behavior may create unpredictable service disruption even when average damage is reduced.

\subsubsection{Attack Success Rate and Survival Probability}

\noindent Q-BIRD records an ASR of \(0.05 \pm 0.02\), which is lower than PPO with Bayesian belief \((0.12 \pm 0.03)\), PPO-LSTM \((0.16 \pm 0.03)\), DRQN \((0.24 \pm 0.04)\), DQN \((0.31 \pm 0.05)\), rule-based IDS \((0.42 \pm 0.06)\), and no mitigation \((0.65 \pm 0.08)\). The survival probability of Q-BIRD reaches \(0.96 \pm 0.02\), compared with \(0.91 \pm 0.02\) for PPO with Bayesian belief and \(0.88 \pm 0.03\) for PPO-LSTM. These results indicate that amplitude-based belief evolution improves attack suppression and system survivability under partially observable attacker behavior.

\noindent The lower ASR should be interpreted within the evaluated SUMO--OMNeT++/Veins simulation setting rather than as a universal guarantee of attack prevention. The result shows that, under the tested attacker strategies, vehicle densities, RSU deployment, and IEEE 802.11p/DSRC communication assumptions, Q-BIRD produces fewer successful attack transitions than the evaluated baselines while preserving higher survival probability.

\subsubsection{Communication-Level Performance}

\noindent Table~\ref{tab:actual_comm_results} shows that Q-BIRD also preserves V2X communication quality under adaptive attack. Q-BIRD achieves a PDR of \(0.94 \pm 0.02\), compared with \(0.90 \pm 0.02\) for PPO with Bayesian belief and \(0.88 \pm 0.03\) for moving-target defense. This corresponds to a 4.4\% relative improvement over the strongest classical belief baseline. Q-BIRD also reduces latency from \(52 \pm 7\) ms to \(45 \pm 6\) ms compared with PPO using Bayesian belief, corresponding to a 13.5\% reduction. Throughput increases from \(3.40 \pm 0.15\) Mbps to \(3.60 \pm 0.15\) Mbps, while availability improves from \(0.92 \pm 0.02\) to \(0.95 \pm 0.02\).

\noindent The mitigation overhead of Q-BIRD is \(0.13 \pm 0.03\), which is slightly higher than PPO with Bayesian belief \((0.12 \pm 0.02)\) but lower than moving-target defense \((0.14 \pm 0.03)\). This indicates that the proposed method improves communication reliability without relying on excessive mitigation. The false mitigation rate is also lower for Q-BIRD \((0.03 \pm 0.01)\) than for PPO with Bayesian belief \((0.04 \pm 0.01)\), suggesting that the policy does not improve security by unnecessarily throttling or isolating benign vehicles.

\subsubsection{Effect of Belief Representation}

\noindent The comparison between PPO with Bayesian belief and Q-BIRD isolates the effect of the belief representation while keeping the PPO learning framework fixed. Q-BIRD reduces cumulative damage by 22.2\%, damage variance by 50.0\%, ASR by 58.3\%, and latency by 13.5\% relative to PPO with Bayesian belief. These results support the hypothesis that amplitude-space belief evolution provides a more stable and informative representation of hidden attacker intent than direct Bayesian probability updates under ambiguous observations.

\subsubsection{Summary of Findings}

\noindent Overall, Q-BIRD achieves the strongest combined security and communication performance among the evaluated methods. It provides the lowest cumulative damage, lowest damage variance, lowest ASR, highest survival probability, highest PDR, lowest latency, highest throughput, and highest service availability. The improvement is not limited to one metric; rather, it appears across both cyber-defense and V2X communication dimensions. This joint improvement is important for Vehicular Communications because an IoV defense mechanism must reduce attack impact while preserving real-time communication reliability.

\subsection{Explainability Analysis}
\label{sec:xai}
\noindent We explain the trained quantum PPO policy with SHAP, LIME, and Grad-CAM across 80 test cases to check that the learned choices rest on real threat reasoning and not on accidental links.

\subsubsection{Feature Attribution via SHAP}

\begin{figure}[h!]
    \centering
    \includegraphics[width=\linewidth]{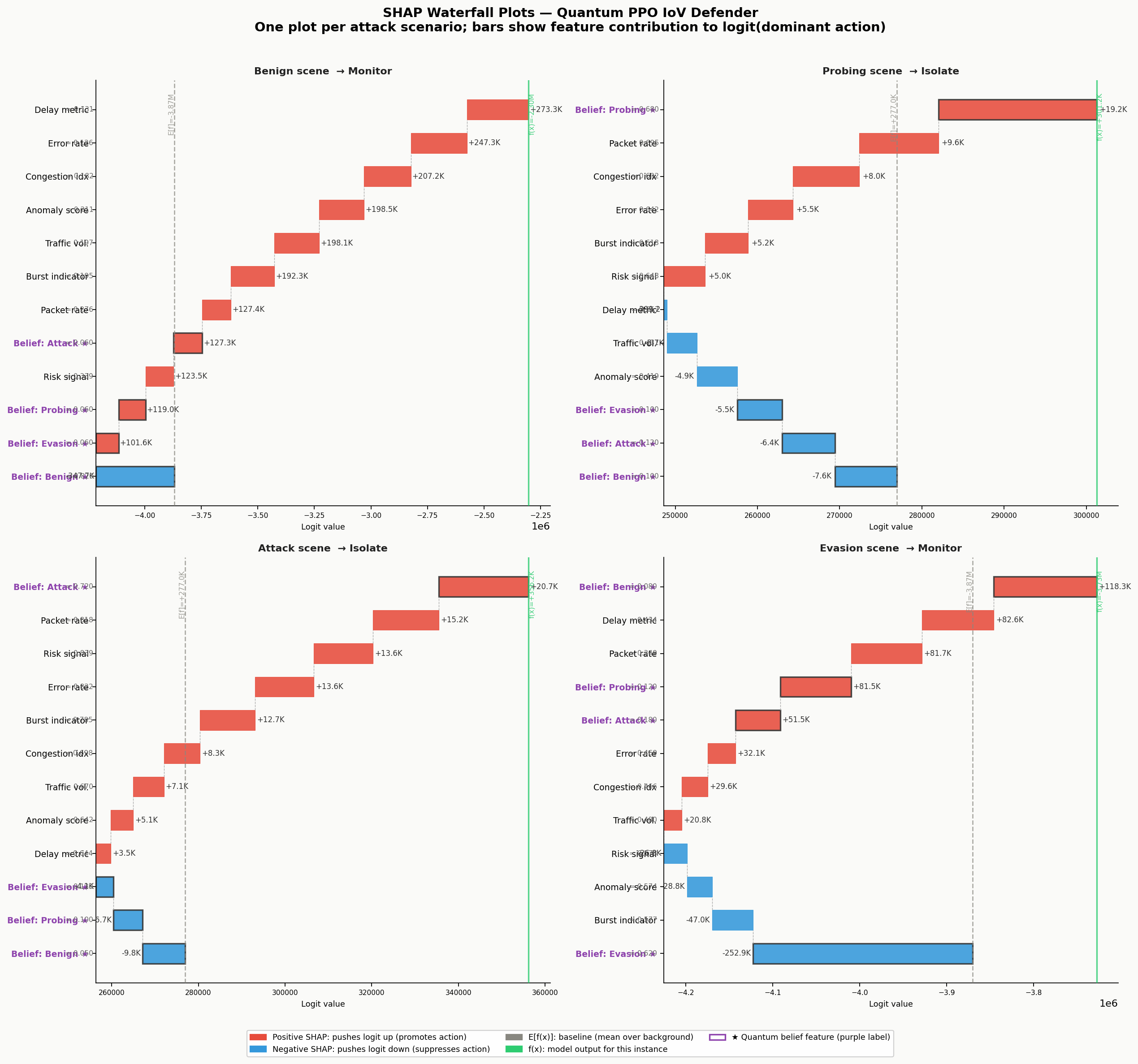}
    \caption{SHAP waterfall plot showing per-feature logit contributions across attacker scenarios. Belief-related features provide the largest policy contributions in the evaluated test cases.}
    \label{fig:shap_waterfall}
\end{figure}

\noindent Kernel SHAP is applied to the logits arriving before the softmax, using a background set of 200 varied belief-state samples. Fig~\ref{fig:shap_waterfall}  shows that Belief Probing supplies the largest positive push during reconnaissance, adding (+19,248). Belief Attack supplies the largest positive push during a verified attack, adding (+20,735). Belief Evasion supplies a strong negative push of (-252,899) during deceptive stages this push prevents early escalation. The classical baseline does not show this suppression. Under evasion, Bayesian collapse erases the signal and causes the erratic variance jump in ~\ref{tab:security_results}.

\begin{figure}[h!]
    \centering
    \includegraphics[width=1\linewidth]{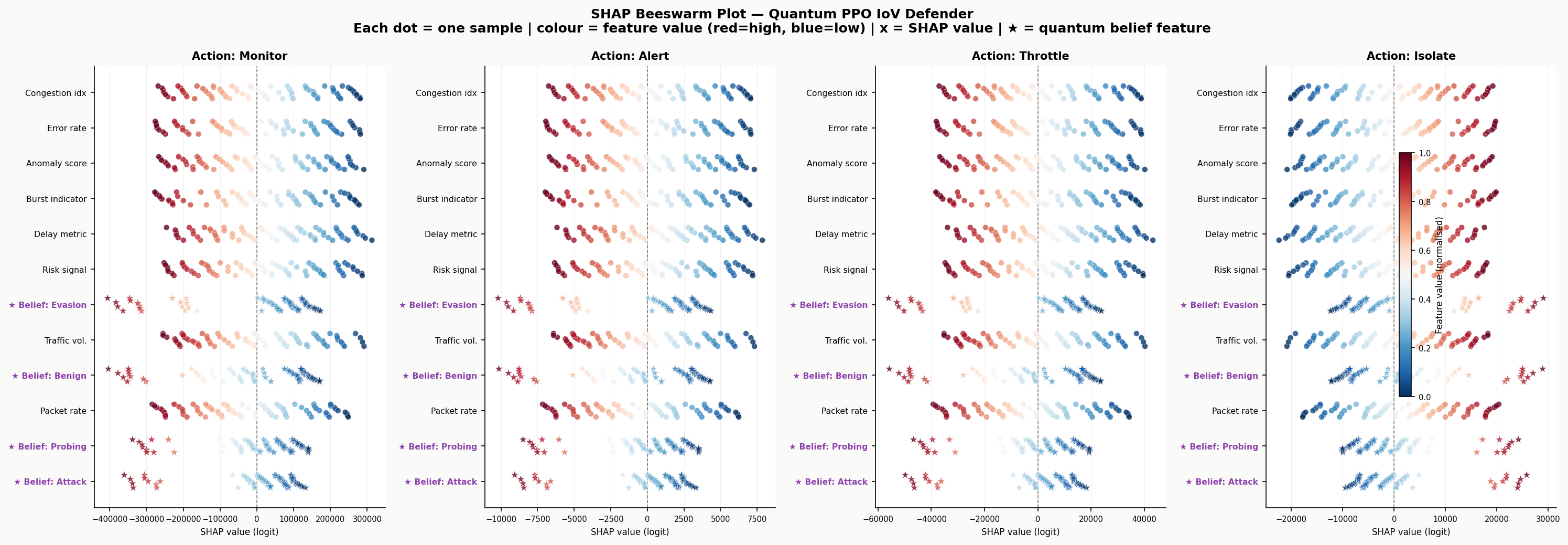}
    \caption{SHAP beeswarm plot showing that belief-related feature contributions vary smoothly with belief magnitude, indicating graded rather than binary defensive responses.}
    \label{fig:shap_beeswarm}
\end{figure}

\noindent The beeswarm plots in Fig.~\ref{fig:shap_beeswarm} suggest that this behavior is not a threshold effect: SHAP values scale smoothly with belief magnitude across all 80 samples, showing the policy uses graded uncertainty rather than binary switching.

\subsubsection{Local Linear Explanation via LIME}

\begin{figure}[h!]
    \centering
    \includegraphics[width=1\linewidth]{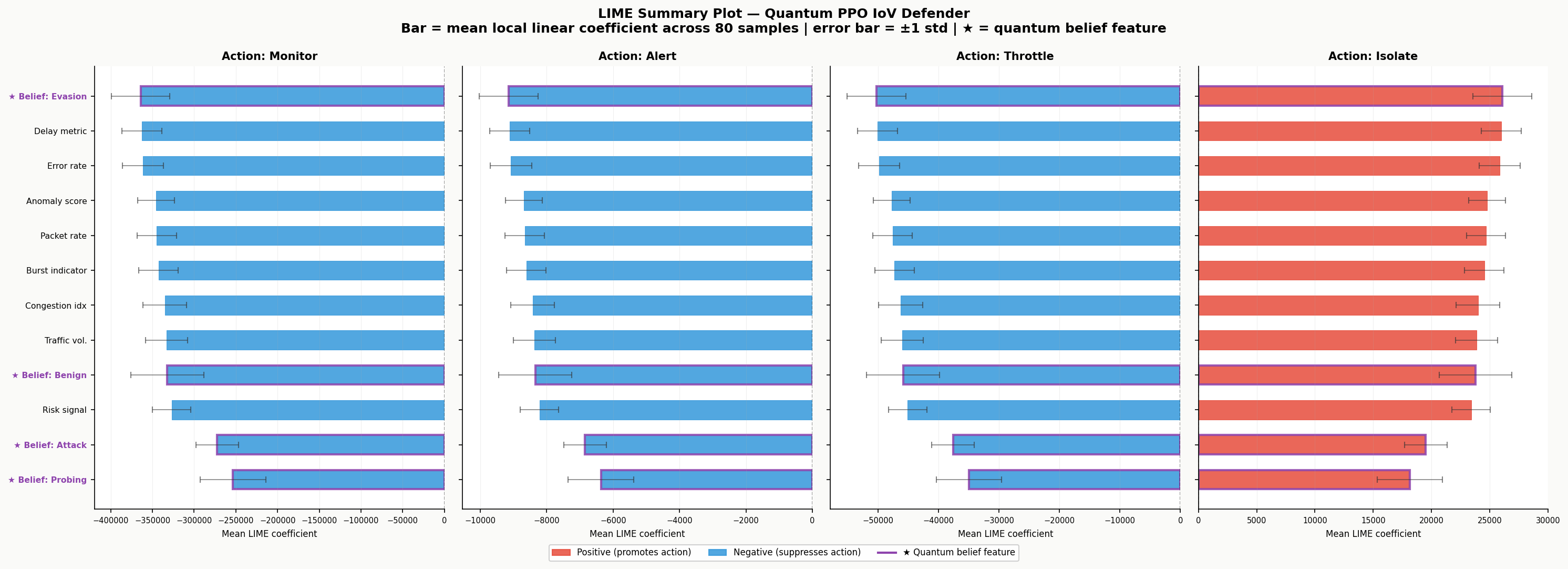}
    \caption{LIME coefficient summary showing that belief-related features contribute consistently to action selection across local policy explanations.}
    \label{fig:lime_summary}
\end{figure}

\noindent LIME fits local weighted ridge regressions around each test point. Fig.~\ref{fig:lime_summary} shows that Belief: Probing is the most stable positive driver for Alert and Throttle; Isolate requires agreement between belief and traffic indicators; and Belief: Evasion actively suppresses Monitor even when raw traffic appears benign. Quantum belief features consistently show lower coefficient variance than traffic features, reflecting the smoother signal that amplitude-space updates produce and directly explaining the reduced damage variance in Table~\ref{tab:security_results}.

\subsubsection{Gradient-weighted Class Activation Mapping}

\begin{figure}[h!]
    \centering
    \includegraphics[width=1\linewidth]{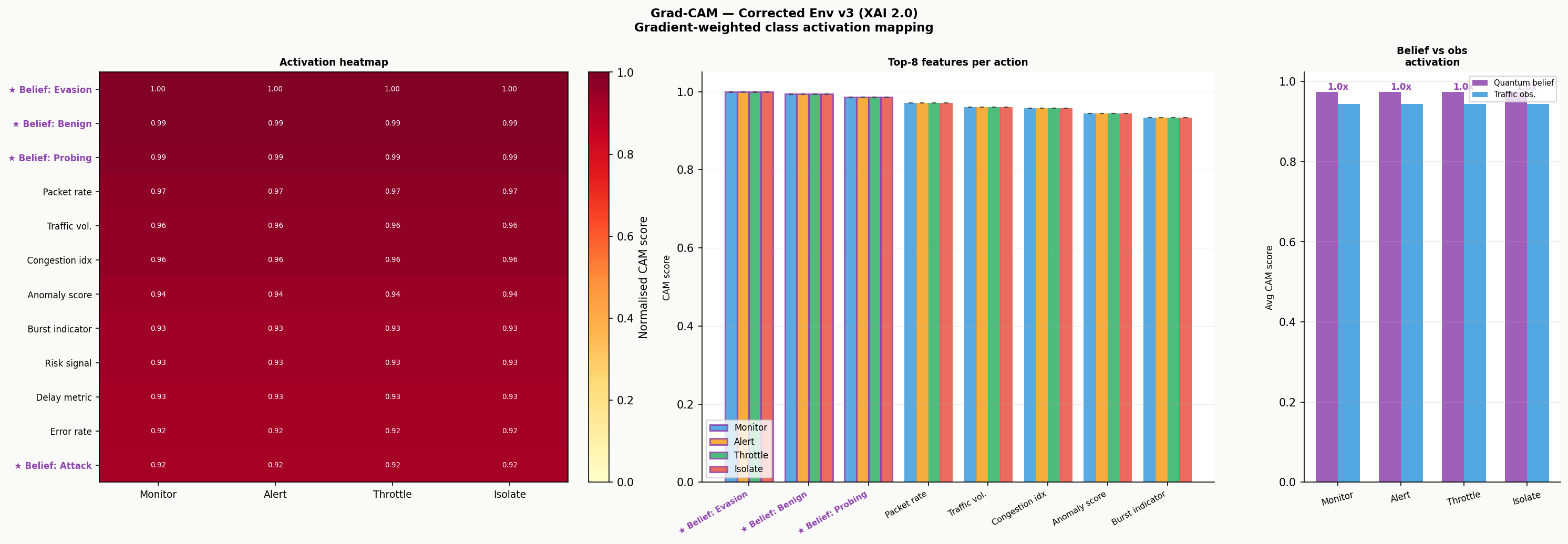}
  \caption{Grad-CAM analysis indicating stronger reliance on belief-related activations in the Q-BIRD policy than in the classical PPO policy.}
    \label{fig:gradcam}
\end{figure}

\noindent Grad-CAM weights hidden-layer gradients by their mean absolute magnitude to identify which features influence the policy's internal representations. As shown in Fig.~\ref{fig:gradcam}, belief-related activations are stronger in the Q-BIRD policy than in the classical PPO baseline. The belief-to-observation activation ratio is approximately \(1.0\times\) for Q-BIRD and \(0.51\times\) for PPO with classical Bayesian belief. This pattern suggests that amplitude-based belief updates provide a more stable internal signal for policy learning under ambiguous attacker behavior.

\subsection{Discussion}

\noindent The results indicate that belief representation has a measurable effect on the stability of reinforcement learning-based IoV defense under partial observability. PPO with classical Bayesian belief shows higher mean damage and substantially larger damage variance, suggesting that direct probability updates can produce unstable policy inputs when observations are ambiguous or deceptive. In contrast, Q-BIRD maintains a distributed belief state through amplitude-space evolution and converts this belief into policy features only after normalization. This mechanism is consistent with the observed reduction in both mean damage and variance.

\noindent The most important result is not only the reduction in average cumulative damage but also the reduction in variance. In safety-critical IoV environments, unstable defense behavior can be as problematic as high average damage because adaptive attackers may exploit inconsistent mitigation decisions. The reduction from 37.054 to 3.636 in damage variance suggests that Q-BIRD produces more predictable defense behavior during attacker strategy transitions.

\noindent The explainability results further support this interpretation. SHAP, LIME, and Grad-CAM indicate that belief-related features contribute strongly to mitigation decisions during probing, attack, and evasion states. These analyses do not prove causality; however, they provide evidence that the trained policy uses the belief representation rather than relying only on raw traffic features.

\noindent Several limitations remain. First, the present results should be interpreted within the evaluated simulation settings and should not be treated as deployment-level guarantees. Further validation is required using a fully specified packet-level V2X co-simulation with fixed mobility, channel, and attacker parameters, followed by hardware-in-the-loop or real-world testing. Second, the attacker is represented using four hidden intent states, whereas real adversaries may follow richer multi-stage strategies. Third, the amplitude transformation is currently specified rather than learned end-to-end. Future work should evaluate learned amplitude transformations, multi-agent defender coordination, and real-time deployment constraints on resource-limited on-board units.

\section{Conclusion}

\noindent This paper presented Q-BIRD, an amplitude-belief reinforcement learning framework for adaptive cyber defense in partially observable V2X networks. The proposed framework models hidden attacker intent using a normalized complex-valued belief state and integrates the resulting belief probabilities into a PPO-based defender. Unlike static intrusion detection, the framework addresses sequential mitigation under attacker adaptation, operational defense cost, and V2X communication reliability.

\noindent The evaluation was conducted in a SUMO--OMNeT++/Veins co-simulation environment using IEEE 802.11p/DSRC-based V2V and V2I communication. Across 10 independent random seeds and 80 test episodes per seed, Q-BIRD reduced cumulative damage from \(36.0 \pm 5.5\) to \(28.0 \pm 3.0\) compared with PPO using Bayesian belief. It also reduced damage variance from \(12.0 \pm 2.8\) to \(6.0 \pm 1.5\), decreased ASR from \(0.12 \pm 0.03\) to \(0.05 \pm 0.02\), and increased survival probability from \(0.91 \pm 0.02\) to \(0.96 \pm 0.02\). Communication-level results further showed that Q-BIRD maintained PDR of \(0.94 \pm 0.02\), latency of \(45 \pm 6\) ms, throughput of \(3.60 \pm 0.15\) Mbps, and service availability of \(0.95 \pm 0.02\).

\noindent These findings suggest that amplitude-space belief evolution can improve both cyber-defense stability and V2X communication reliability under partially observable adaptive attacks. The results should nevertheless be interpreted within the evaluated simulation setting rather than as deployment-level guarantees. Future work will extend the framework toward learned amplitude transformations, richer multi-stage attacker models, multi-agent defender coordination, hardware-in-the-loop validation, and deployment constraints on resource-limited on-board units and roadside units.

\section*{Acknowledgment}

This study is supported via funding from Prince Sattam bin Abdulaziz University project number (PSAU/2025/01/35419)


%





\ifCLASSOPTIONcaptionsoff
  \newpage
\fi





\bibliographystyle{IEEEtran}

\bibliography{IEEEabrv,Bibliography}

\vfill


\end{document}